# On the kinematic and thermodynamic state of clouds in complex wind–multicloud environments using a friends-of-friends analysis

A. Antipov,[1★] W. E. Banda-Barragán ,[2,3] Y. Birnboim ,[1] C. Federrath ,[4,5] O. Gnat[1] and M. Brüggen [3]

[1] *Centre for Astrophysics and Planetary Science, Racah Institute of Physics, The Hebrew University, Jerusalem 91904, Israel*
[2] *Escuela de Ciencias Físicas y Nanotecnología, Universidad Yachay Tech, Hacienda San José S/N, 100119 Urcuquí, Ecuador*
[3] *Hamburger Sternwarte, University of Hamburg, Gojenbergsweg 112, D-21029 Hamburg, Germany*
[4] *Research School of Astronomy and Astrophysics, Australian National University, Canberra, ACT 2611, Australia*
[5] *Australian Research Council Centre of Excellence in All Sky Astrophysics (ASTRO3D), Canberra, ACT 2611, Australia*

## ABSTRACT

We investigate the interaction between a shock-driven hot wind and a cold multicloud layer, for conditions commonly found in interstellar and circumgalactic gas. We present a method for identifying distinct clouds using a friends-of-friends algorithm. This approach unveils novel detailed information about individual clouds and their collective behaviour. By tracing the evolution of individual clouds, our method provides comprehensive descriptions of cloud morphology, including measures of the elongation and fractal dimension. Combining the kinematics and morphology of clouds, we refine previous models for drag and entrainment processes. Our by-cloud analysis allows to discern the dominant entrainment processes at different times. We find that after the initial shock passage, momentum transfer due to condensation becomes increasingly important, compared to ram pressure, which dominates at early times. We also find that internal motions within clouds act as an effective dynamic pressure that exceeds the thermal pressure by an order of magnitude. Our analysis shows how the highly efficient cooling of the warm mixed gas at temperatures $\sim 10^5$ K is effectively balanced by the kinetic energy injected by the hot wind into the warm and cold phases via shocks and shear motions. Compression-driven condensation and turbulence dissipation maintain a multiphase outflow and can help explain the presence of dense gas in galaxy-scale winds. Finally, we show that applying our friends-of-friends analysis to H I-emitting gas and correcting for beam size and telescope sensitivity can explain two populations of H I clouds within the Milky-Way nuclear wind as structures pertaining to the same outflow.

## 1 INTRODUCTION

The circumgalactic medium (CGM) is a large reservoir of gas that surrounds galaxies. Observational studies (e.g. see Rupke 2018) and theoretical work (e.g. see Birnboim & Dekel 2003) show that the CGM and the outflows within it are multiphased media, containing dust (Kane & Veilleux 2024), 'cold' $< 10^3$ K molecular gas (Di Teodoro et al. 2020), 'warm' $\sim 10^4$ K atomic gas (Ciampa et al. 2021), 'hot' $\sim 10^6$ K ionized gas (Das, Mathur & Gupta 2020), and even a 'very hot' $\sim 10^7$ K component (Gupta et al. 2021) within them. The cold gas phase is of interest because its creation and survival within the background, low-density, hot gas for extended periods of time presents a theoretical challenge (Zhang 2018). Transporting or producing cold gas to/at large radial distances in the CGM requires galactic winds, which can arise from supernovae (SNe) feedback (e.g. see Llerena et al. 2023; Perrotta et al. 2023) or active galactic nuclei (AGNs) feedback (e.g. see Harrison & Ramos Almeida 2024; Speranza et al. 2024). In either case, the interstellar medium (ISM) plays a crucial role as it is the source of mass loading and entrainment in galactic winds (Zhang et al. 2021). The ISM is turbulent and multiphased, so studying how winds are launched and evolve in such environments (and at different scales) is crucial in order to understand the baryon cycle in galaxy evolution (Tumlinson, Peeples & Werk 2017).

Galactic outflows have been observed in the Milky Way (MW; Su, Slatyer & Finkbeiner 2010) as well as in other galaxies (Shopbell & Bland-Hawthorn 1998). Observations of the nuclear outflow in our Galaxy are particularly important because gas phases can be seen in greater detail than in external galactic winds. Such observations confirm the multiphase nature of outflows, with gas phases observed at a wide range of temperatures. X-ray observations reveal a hot and diffuse component (Predehl et al. 2020), while emission- and absorption-line studies of different atomic and molecular species show the presence of a $10^3 - 10^4$ K atomic gas phase (Lockman, Di Teodoro & McClure-Griffiths 2020), and a $< 10^2$ K molecular gas phase (Veena et al. 2023; Noon et al. 2023). The cold gas phase is observed in compact clouds, moving at high local standard of rest (LSR) velocities $\gtrsim 200$ km s$^{-1}$ and at substantial distances above and below the galactic plane $\sim 1$ kpc (McClure-Griffiths et al. 2013; Di Teodoro et al. 2018), suggesting a scenario of dense gas entrainment within the hot and diffuse wind.

★ E-mail: andrei.antipov@mail.huji.ac.il

The acceleration of cold clouds within a hot wind, or as it is also often called the 'cloud-crushing problem', is a non-trivial process. The slow(er) moving cold cloud in the presence of a fast, hot wind develops Kelvin–Helmholtz (KH) instabilities at the interface between the two phases. These shear instabilities can drive turbulence over a cascade of scales (Mandelker et al. 2016), eventually mixing the cold gas with the hot wind (Fielding et al. 2020), to a state where the initial cold cloud is indiscernible from the hot wind. In the purely hydrodynamical case, Klein, McKee & Colella (1994) showed that a single cold cloud is eventually broken up and mixed with the hot wind, over only a few cloud-crushing times, $t_{\rm cc} = \chi^{1/2} r_{\rm cl}/u$, where $\chi = \rho_{\rm cl}/\rho_{\rm w}$ is the density contrast between the cold cloud and the wind, $u$ is the relative velocity between them, and $r_{\rm cl}$ is the initial cloud radius (some authors prefer to use other length-scales, e.g. the cloud diameter is used in Jones, Ryu & Tregillis 1996). For density contrasts commonly observed in the CGM, $\chi \gtrsim 100$, so the typical cold cloud entrainment time-scale, $t_{\rm ent} \sim \chi r_{\rm cl}/u \propto \chi^{1/2} t_{\rm cc}$, exceeds several cloud-crushing times. The cold phase in adiabatic models is then readily disrupted by the hot wind before it can become entrained within it. To explain why we detect cold phases in observations, (re)condensation (Marinacci et al. 2010), magnetic fields (Cottle et al. 2020), and hydrodynamic shielding (Villares, Banda-Barragán & Rojas 2024) have been shown to sustain dense gas.

Once radiative cooling is added to the scenario, with a cooling function $\Lambda(T, n)$, for temperature $T$, and number density $n$, there exists a span of physical parameters that allows the cold gas to survive for extended periods of time (Gronke & Oh 2018), and even increase in its volume and mass (Gronke & Oh 2020; Sparre, Pfrommer & Ehlert 2020; Kanjilal, Dutta & Sharma 2021). Once a parcel of cold material mixes with hot gas, an intermediate 'warm' gas phase emerges, at a typical mixing temperature $T_{\rm mix} \sim \sqrt{T_{\rm cold} T_{\rm hot}}$ (Begelman & Fabian 1990). For the conditions relevant to gas pulled from the ISM into the CGM, $T_{\rm mix} \sim 10^5$ K. At this mixing temperature, the cooling rate increases by $\sim 2$ orders of magnitude (Gnat & Sternberg 2007; Teşileanu, Mignone & Massaglia 2008), so warm mixed gas cools down very efficiently. This condensation process allows for warm-mixed gas to cool down to $\sim 10^4$ K, which is commonly used as the floor temperature for radiative cooling ($\sim 10^4$ K is the temperature at which the heating by a background ultraviolet (UV) radiation balances radiative cooling).

The gas cooling time is defined as $t_{\rm cool} \propto T/n\Lambda(T, n)$. Once a system is set up with the conditions where the cloud-crushing time is longer than the cooling time-scale, $t_{\rm cool}/t_{\rm cc} < 1$, the dense gas phase is expected to survive the disruptive forces acted on it by the hot wind, until it is finally entrained in the wind. The cooling time and the gas sound speed, $c_{\rm s}$, defines a typical length-scale for the cold material, $r_{\rm cloudlet} = c_{\rm s} t_{\rm cool}$. In setups where the cooling time is substantially smaller than the cloud-crushing time, $t_{\rm cool}/t_{\rm cc} \ll 1$, the cloud is expected to cool rapidly to small cloudlets of a typical length-scale, $r_{\rm cloudlet}$. In cases where the initial cloud size is large with respect to the cloudlet radius, $r_{\rm cl,\,init}/r_{\rm cloudlet} \gg 1$, the initial cold cloud 'shatters' into numerous separate cold cloudlets of size $\sim c_{\rm s} t_{\rm cool}$ (McCourt et al. 2018).

This complex behaviour of cold gas embedded in a fast-moving hot wind has been extensively researched in simulations at multiple ISM and CGM scales. At small scales, a single isolated cloud is placed in 'wind tunnel' setups (Cooper et al. 2009; Gronke & Oh 2018; Abruzzo, Bryan & Fielding 2022). These models capture cloud disruption, mixing processes, and condensation at high resolutions (Cooper et al. 2009). At medium scales, 'tall box' simulations capture wind launching and intracloud interactions (Banda-Barragán et al. 2020; Kim et al. 2023). At large scales, the disc- and stellar-driven outflows are captured (Cooper et al. 2008; Sarkar, Nath & Sharma 2015; Schneider et al. 2020), and there are also zoom-in CGM simulations taken from large-scale cosmological simulations (Hummels et al. 2019; Nelson et al. 2020), which cover the full extent of the wind. The level of physical complexity varies between these simulations. It has been shown that the addition of thermal conduction (Gnat, Sternberg & McKee 2010; Brüggen & Scannapieco 2016), magnetic fields (Grønnow, Tepper-García & Bland-Hawthorn 2018; Sparre et al. 2020; Casavecchia et al. 2024), turbulence (Banda-Barragán et al. 2019; Fournier et al. 2024), and cosmic rays (Wiener, Zweibel & Ruszkowski 2019; Brüggen & Scannapieco 2020) can change the morphology and entrainment evolution of the cold clouds. For instance, thermal conduction compresses clouds with high-column densities into filaments, magnetic fields transverse to the wind can reduce KH instabilities, supersonic turbulence can contribute to cloud disruption, and cosmic rays embedded in magnetic fields aligned with the wind can pile up in front of clouds and accelerate them.

In this paper, we use medium-scale, radiative wind–multicloud simulations (see Banda-Barragán et al. 2021) to study the individual and collective properties of clouds within hot outflows using a friends-of-friends (FoF) clustering algorithm. These simulations are appropriate for our study because they have a turbulent initial density distribution and sufficient resolution to capture the cold phases. Besides, the complex initial spatial distribution of the cold phase follows a lognormal density distribution and evolves into a multiphase outflow with an ensemble of cold droplets, which are the targets of our study.

The structure of the paper is as follows. In Section 2, we motivate our study. In Section 3, we present the wind–multicloud set of simulations. In Section 4, we present our FoF algorithm, and show some of its results, evolution of clouds' number, masses, locations, and velocities. In Section 5, we discuss the morphology and dynamics of the clouds. Section 6 presents the analysis of the thermodynamical state of the clouds. In Section 7, we compare our results to observations of cold clouds in the MW, and in Section 8, we summarize the results.

## 2 SIGNIFICANCE OF TRACKING THE COLD GAS

Tracking dense gas in simulations can be done in several manners and is an important step towards developing a coherent picture of how the cold phases evolve in galactic winds. Often, the analysis of the cold phase and its evolution is performed by summing or averaging over the entire cold gas in the computational domain. This is done by counting all the gas cells which lie below some physically motivated temperature threshold, or above a set density value. Another way to follow the cold phase is to plant scalar fields of tracer particles in the initial setup (e.g. Federrath et al. 2008a; Banda-Barragán et al. 2019), and then follow the evolution of that scalar field. This scalar field can then be used as a weighting function to average only over the gas initially associated with the cold phase. Such analyses on the scale of the entire cold phase present limitations as a wealth of small-scale information is hidden in the evolution of individual clouds.

As wind–cloud (e.g. Gronke & Oh 2018) and wind–multicloud (e.g. Banda-Barragán et al. 2021) simulations have demonstrated, dense gas in multiphase outflows is often the result of cooling-driven condensation from the warm mixed gas or from the hot wind. Cold droplets may also be the result of the break-up of initially larger cold clouds (McCourt et al. 2018) or coalescence mechanisms (Waters & Proga 2019). In general, clouds constantly change their morphology and surface due to their interaction with the hot wind, other clouds, and the intercloud material. To track such changes, a more detailed analysis with clump/cloud recognition routines is required.

A large variety of clump recognition algorithms exists. To name a few: the classical FoF type of search (Davis et al. 1985) is considered computationally efficient, but it tends to overconnect separate dense objects via narrow 'bridges' of dense gas. This is referred to as the 'linking length'. Eisenstein & Hut (1998) present a more accurate and computationally expensive hopping algorithm ('HOP'). This method operates on particles by assigning each one a density value and moving up the density gradient through a series of 'hops' between particle pairs. At each hop, the algorithm moves from a lower density neighbour to a higher density one until it reaches a particle that is its own densest neighbour – hence the name. Ultimately, every particle in the data set is assigned to a single, distinct group. We note that in mesh-based simulations, like those we present in this paper, the algorithm works on grid cells, rather than particles. This algorithm also uses a series of density thresholds which reduce the overconnectedness of clumps via linking lengths. An additional improvement was made by Skory et al. (2010) via the introduction of the 'Parallel HOP' method to reduce the computational cost. These methods (and more) are easily accessible on the web, e.g. see the 'YT' software package (Turk et al. 2011).

The strength of a 'by-cloud' analysis of simulations is demonstrated well in numerous papers, as they uncover information untapped by a 'phase-scale' analysis. Gronke & Oh (2023) use the evolution of the amount of separate cold clouds to measure their rate of coagulation. Similarly, Sparre, Pfrommer & Vogelsberger (2019) measure the number of separate clouds to trace the fragmentation of a cloud using FoF. Tonnesen & Bryan (2021) use the results of a cloud recognition routine, combined with an analysis of a tracer scalar field, to discern the regime of entrainment of a satellite galaxy by the CGM. Tan & Fielding (2024) use an FoF analysis to study the evolution of the surface area of clouds, their intrinsic velocities, the rate of inflow of new condensed material, and the properties of turbulence.

In this paper, we introduce and apply our own FoF algorithm (coded in PYTHON) on two simulations of a supersonic hot wind passing through a cold multicloud layer, in the presence of radiative cooling and heating by a background radiation field. We use our FoF analysis on a simulation presented in Banda-Barragán et al. (2021) to extract new results from it accessible only with our by-cloud FoF analysis. Additionally, we report a new simulation with the same computational setup, but using a lower resolution, which allows us to follow the evolution of the multicloud system for a longer time and a larger streaming distance. We examine the evolution of the resulting multiphase gas in a phase-scale and a by-cloud type of analysis.

## 3 WIND–MULTICLOUD SIMULATIONS

### 3.1 Simulation description

In order to study the properties of individual clumps in multiphase CGM environments, we resort to numerical simulations of wind–multicloud systems. Here, we provide a brief summary of the setups of these simulations. The base simulations for our analysis were originally reported in Banda-Barragán et al. (2020, 2021) and consist of 3D models of hot, supersonic shocks interacting with cold, multicloud systems. In this paper, we analyse two Eulerian simulations, performed using the PLUTO v4.3 code (Mignone et al. 2007) with uniform grids:

(i) High-resolution model ('high-res'), which corresponds to a simulation reported in Banda-Barragán et al. (2021), specifically to model sole-k8-M10-rad (see table 1 in that paper). The domain size of this simulation is $100 \times 500 \times 100\,\mathrm{pc}^3$, with a numerical resolution of $0.39^3\,\mathrm{pc}^3$.

(ii) Low-resolution model ('low-res'), which corresponds to a new simulation, based on the same above model (high-res), but specifically ran for this study in a larger computational domain. The volume of this simulation is $100 \times 1500 \times 100\,\mathrm{pc}^3$, with a numerical resolution of $0.78^3\,\mathrm{pc}^3$. As we discuss in Section 4 and thereafter, this choice of domain setup allows us to simultaneously study the late-stage evolution of multiphase outflows and the resolution dependence in our FoF clump analysis (while optimizing computational cost).

Our wind–multicloud simulations are initialized with three components at $t = 0$ (see the leftmost column of Fig. 1):

(i) *A pre-shock region*. The pre-shock ambient medium (in light pink colour) has a constant number density of $n_{\mathrm{ambient}} = \rho_{\mathrm{ambient}}/\mu\, m_u = 0.01\,\mathrm{cm}^{-3}$, where $\mu$ is the mean atomic weight per particle, $m_u$ is the atomic mass unit, a constant thermal pressure of $p_{\mathrm{ambient}}/k_{\mathrm{b}} = 10^4\,\mathrm{K\,cm}^{-3}$, and a temperature of $T_{\mathrm{ambient}} = 10^6\,\mathrm{K}$.

(ii) *A post-shock flow (the wind)*. The post-shock region (in dark red colour) is created by a planar shock wave, driven along the vertical axis of the simulated domain. The shock is driven by imposing a supersonic inflow of gas through the bottom boundary of the domain. The shock speed is set to $v_{\mathrm{shock}} = 1440\,\mathrm{km\,s}^{-1}$, which corresponds to a Mach number of $\mathcal{M} = v_{\mathrm{shock}}/c_{\mathrm{ambient}} = 10$, for $c_{\mathrm{ambient}} = 144\,\mathrm{km\,s}^{-1}$. The post-shock ambient medium has higher density, pressure, and temperature than the pre-shock gas and is set up according to the Rankine–Hugoniot jump conditions (Landau & Lifshitz 1987). Thus, its speed is $v_{\mathrm{wind}} = 1080\,\mathrm{km\,s}^{-1}$, its number density is $n_{\mathrm{psh}} \approx 4\,n_{\mathrm{ambient}}$, its thermal pressure is $p_{\mathrm{psh}} \approx 125\,p_{\mathrm{ambient}} = 1.3 \times 10^6\,k_{\mathrm{B}}\,\mathrm{K\,cm}^{-3}$, and its resulting temperature is $T_{\mathrm{psh}} = 3 \times 10^7\,\mathrm{K}$.

(iii) *A multicloud layer*. The multicloud layer (in light blue colour) is positioned on the lower side of the domain, covering the full horizontal width of the 3D domain, and contains a turbulent density field, motivated by density fields observed in steady-state turbulence simulations. The lognormal density field for the multicloud layer is generated with the PYFC library (available at: https://bitbucket.org/pandante/pyfc), which constructs randomly generated, periodic scalar fields that follow pre-defined power-law spectra, $D(k) \propto k^{-0.78}$. This exponent for the density power spectrum was chosen as it is the result of purely solenoidal driving of turbulence in a supersonic medium (Federrath, Klessen & Schmidt 2009), which serves a reasonable density distribution for the numerical experiments carried out here. The density fluctuations of this turbulent field were also chosen to agree with this (Federrath et al. 2010). We leave a study that starts with a purely compressively driven turbulent initial condition (Federrath, Klessen & Schmidt 2008b) for a future study, and refer to Banda-Barragán et al. (2020, 2021) for a comparison. The observed clumpiness of the layer is controlled by the minimum wavenumber set for such a power law, which for these models is $k_{\mathrm{min}} = 8$ (for further details, the reader is referred to section 2.3 in Banda-Barragán et al. 2020). The multicloud layer gas has an initial mean number density value of $\bar{n}_{\mathrm{cloud},0} = 1\,\mathrm{cm}^{-3}$ (so it is on average a 100 times denser than the pre-shock region) and it is initially held in pressure equilibrium with the pre-shocked ambient medium with zero velocity.

Both simulations were initialized with the same initial conditions, were seeded with the same initial multicloud layer, and were evolved in time in the rest frame of the pre-shocked medium. The simulated medium is a mono-atomic ideal gas, $\gamma = 5/3$, without gravity present. The simulation code includes both radiative cooling by atomic coolants and radiative heating by a background UV field, in the temperature range between $10^2$ and $10^7$ K, which correspond to our numerical cooling/heating floor and ceiling, respectively. The net cooling and heating function depends on both temperature and density (see Fig. 2). The cooling function (computed with CLOUDY, Ferland et al. 1998) corresponds to atomic gas with Solar metallicity, and the heating function mimics the redshift-zero metagalactic Haardt and Madau (2012) UV background. The heating function is normalized such that at $T = 10^4$ K and $n = 10^{-1}$ cm$^{-3}$, heating balances cooling. For relatively low densities, the cooling function significantly drops below $T \sim 10^4$ K, so below this temperature, cooling drops, and heating becomes important. On the other hand, for relatively high densities, the gas can reach thermal equilibrium at lower temperatures, since heating that scales with the density – $n$, is less effective than cooling that scales with – $n^2$. This implies that some high-density gas in the clouds can cool down to temperatures of $\sim 10^2$ K (our choice of cooling floor). For a full and detailed discussion of the numerical setup, we refer the reader to section 2 of Banda-Barragán et al. (2021).

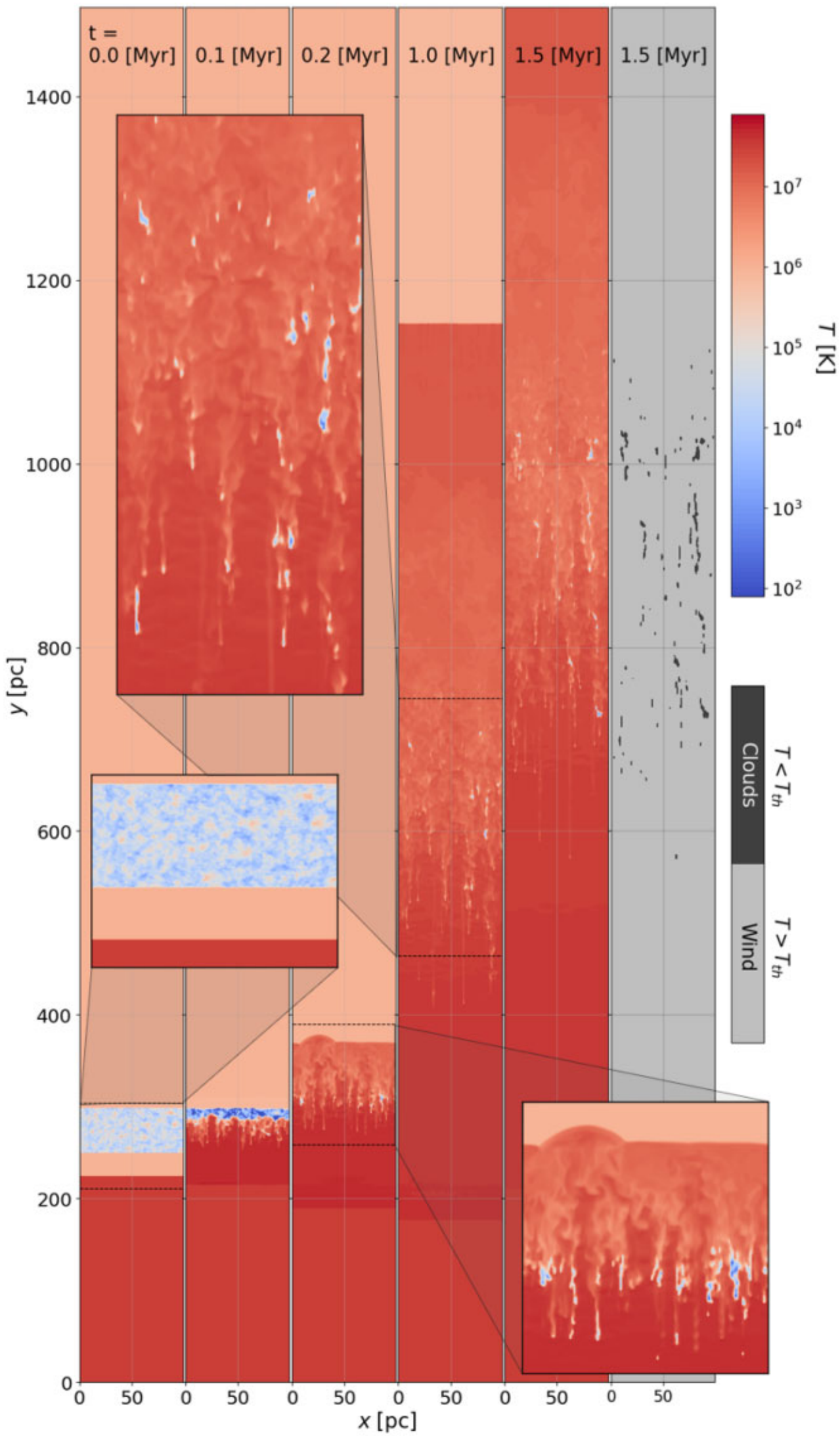

**Figure 1.** Visualization of the temperature distribution in the low-res simulation at a few selected times. The five leftmost panels present $\hat{x} - \hat{y}$ slices along the centre of the box in the $\hat{z}$-direction. Dark red areas correspond to the post-shock flow (i.e. to shocked gas, $T > 10^7$ K, which we refer to as the 'wind'), pink areas to the pre-shock material ($\sim 10^6$ K), while the blue and intermediate colours correspond to cold gas in the multicloud layer with a lognormal density distribution ($\lesssim 10^5$ K). Zoom-in bubbles highlight the initial shock-multicloud setup and the resulting post-shock flow. The right panel shows the input to the FoF analysis, where the black cells meet the threshold temperature criterion $T < T_{\rm th}$ ($= 10^{5.5}$ K) and are considered to be cloud cells, while the grey cells are treated by the algorithm as non-cloud gas.

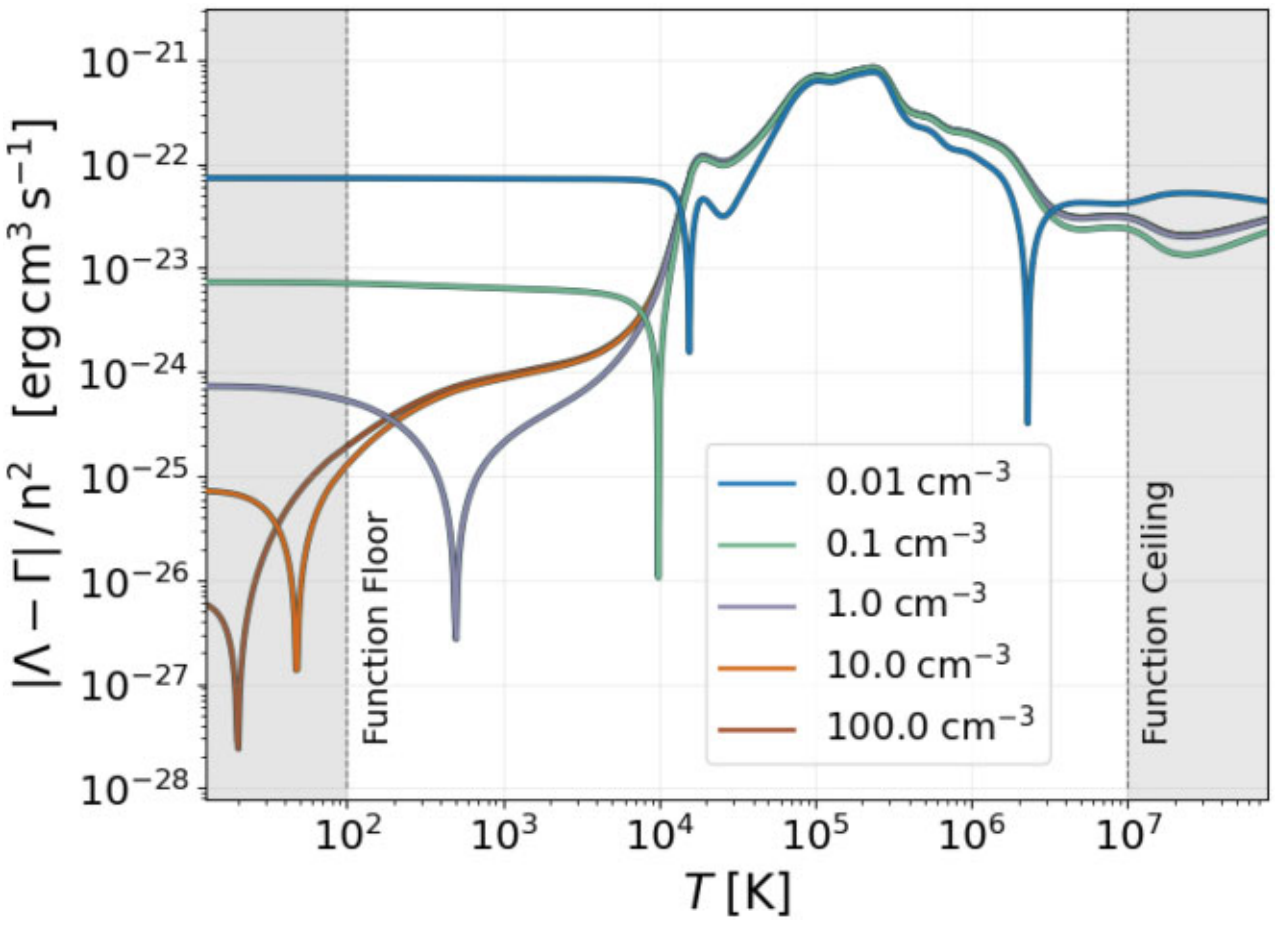

**Figure 2.** Net cooling–heating function used in our set of simulations. The lines show cooling/heating rates for several values of gas number density, with the dashed vertical lines at $10^2$ and $10^7$ K showing the numerical cooling/heating floor and ceiling we used.

### 3.2 Evolution of wind–multicloud systems

The evolution of our wind–multicloud system (in model low-res), at different stages of the run, is presented in Fig. 1. Each slice in this figure represents a state of the system at increasing times (from left to right). At $t = 0$, the very hot material at the bottom is the post-shock flow (i.e. the wind), the gas permeating most of the volume is the uniform pre-shock region, and the cold material (residing within it) is associated with the initial multicloud (turbulent) layer. The panels at later times show the aftermath of the interaction between the shock (and the post-shock flow or wind) and the clouds. The evolution of this model is akin to the one described for model high-res (i.e. model sole-k8-rad in Banda-Barragán et al. 2021) and also confirms the results presented in Alūzas et al. (2012) and Banda-Barragán et al. (2020), i.e. that radiative wind–multicloud systems evolve in four stages: (1) the shock first interacts with the multicloud layer, transmitting internal shocks that start heating up the gas in it, (2) the transmitted shocks travel through the multicloud layer compressing the gas and then emerge on the downstream side at $\sim 0.17$ Myr, (3) the transmitted shocks re-accelerates into the downstream (more diffuse) pre-shock gas triggering filament formation and the onset of KH instabilities, and (4) the interaction between the wind and the shocked multicloud gas leads to mixing, turbulence, and the formation of a multiphase flow characterized by a rain-like morphology.[1]

[1]For a larger collection of simulation images and video clips, we refer the reader to: https://andreiantipov.huji.ac.il/.

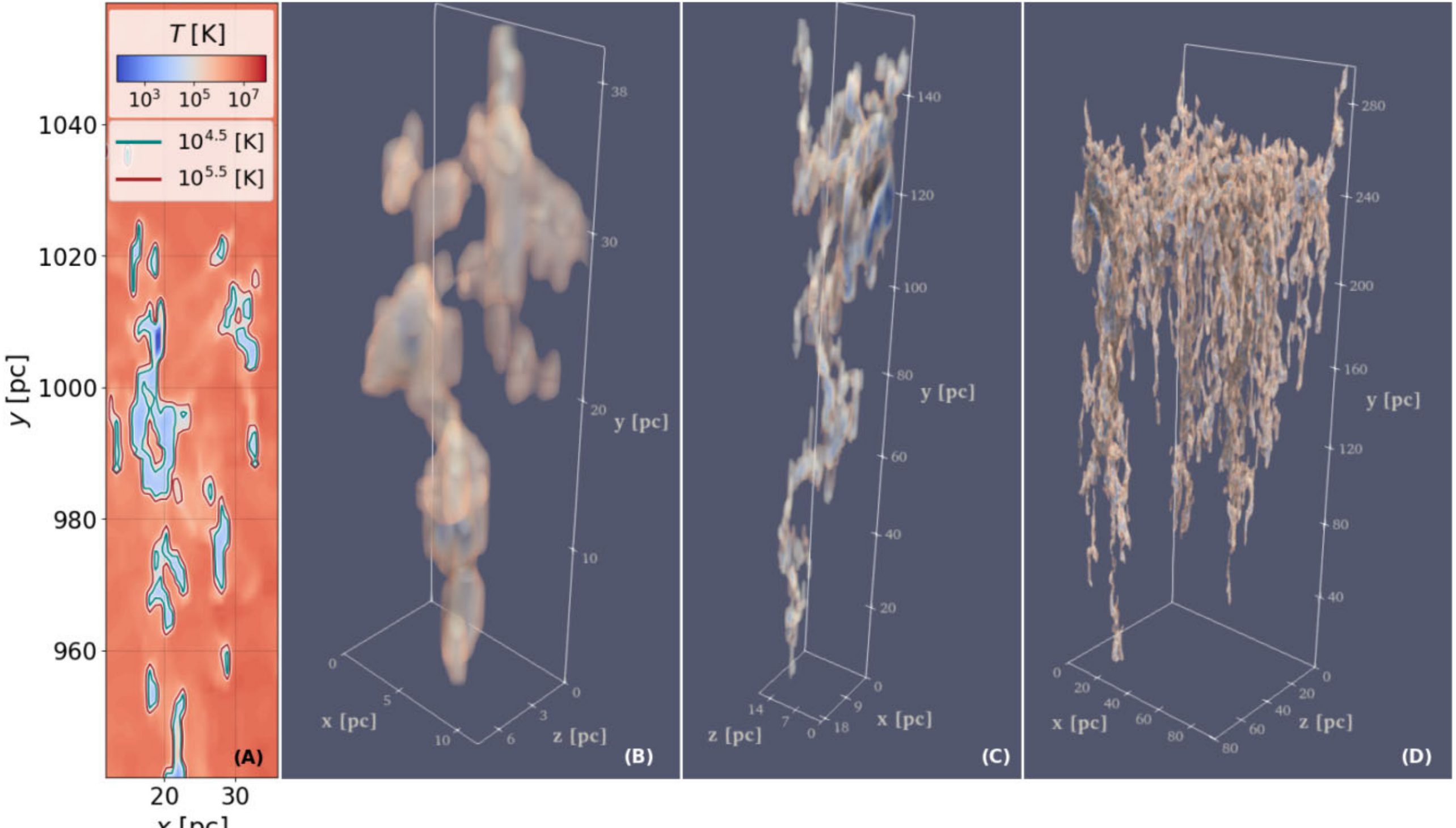

**Figure 3.** (a) Zoom-in image into a 2D slice of the temperature distribution in the low-res simulation at $t = 1.5$ Myr. Contour lines denote the iso-surface line for $T = 10^{4.5}$, and $10^{5.5}$ K, which depict the influence of the choice of threshold temperature on the recognition and merger of cold clouds. Then, progressing to the right, also for $t = 1.5$ Myr, we present 3D renderings of the results of our FoF analysis for $T_{\rm th} = 10^{5.5}$ K. FoF outputs of clouds of different sizes: (b) 691 cells $\simeq$ 328 pc$^3$, (c) 6455 cells $\simeq$ 3060 pc$^3$, with the largest recognized cloud in this snapshot depicted in (d) 87497 cells $\simeq$ 41 520 pc$^3$. Note the different scale size in all three renderings. In panels (b)–(d), all gas cells $T > 10^{5.5}$ were set to be transparent. The 3D cloud renderings share the same logarithmic colour-scheme, shown in the colour bar of image (a).

While radiative cooling and phase separation are initially enhanced by the abrupt contraction and heating driven by internal shocks, the subsequent onset of turbulence (for $t > 0.2$ Myr), which occurs once the main shock has fully passed through the initial vertical length of the multicloud gas leads to the formation of a stable multiphase flow with three distinct phases: a cold one at $\sim 10^2$ K, a warm one at $\sim 10^4$ K, and hot one at $10^7$ K (see section 3.2 in Banda-Barragán et al. 2021 for additional details). The cold and warm phases are spatially coexistent and comprise dense filaments with low volume filling factors, whilst the hot phase is pervasive. Dense gas in the evolving multiphase flow is not directly entrained by the post-shock flow, but it is rather destroyed by KH and Rayleigh–Taylor (RT) instabilities (in agreement with previous wind-single cloud studies by Cooper et al. 2009; Schneider & Robertson 2017; Banda-Barragán et al. 2019) and subsequently reformed via condensation driven by radiative cooling (in agreement with prior work by Gronke & Oh 2020; Li et al. 2020).

## 4 FRIENDS-OF-FRIENDS ANALYSIS

### 4.1 Temperature threshold and cold gas properties

We wish to identify the cold-phase clouds based on their thermodynamic properties. This requires setting up a physically defined threshold on one of the gas properties. A variety of physical properties and threshold values can be used, e.g. a number density threshold ($n_{\rm FoF,th}$ as in Sparre et al. 2019), or a specific ion ($Mg_{\rm II}$ as in Nelson et al. 2020). The threshold can also be applied directly to the temperature field, such as $\lesssim 10^{4.5}$ K in Tonnesen & Bryan (2021) or $< 10^{5.5}$ K as in Tan & Fielding (2024). In this work, we consider a rolling temperature threshold, which allows us to progressively peel-off hot layers of clouds, while tracing how properties of individual clouds change. Fig. 3 presents the results from running our FoF analysis on our low-res wind–multicloud simulations. Particularly, we carried out our analysis using three different threshold values: the 'cold' phase was defined as the cells that lie below the threshold temperature of $T_{\rm th}[{\rm K}] \in \{10^{4.5}, 10^{5.0}, 10^{5.5}\}$. All results presented here were checked for these three threshold values. We have verified that the qualitative results presented below are independent of the choice of $T_{\rm th}$.

Fig. 4 presents the time evolution of the ensemble of cold clouds in our simulations. The upper three panels of Fig. 4 show the time evolution of the total mass ($M_{\rm tot}$), velocity ($v_{\rm y,tot}$), and vertical displacement ($y_{\rm tot}$) of the cold gas. We show the results for our three values of the threshold temperature (in different colours), and for two resolutions (the solid line is used for the low-res model and the dashed line for the high-res one). The mass is calculated by integrating the cell densities over their respective volumes, while the velocity and displacement of the centre of mass are mass-weighted averages. These three properties show a smooth and monotonically growing evolution, regardless of the chosen threshold. Mass grows as new material accretes to the cold phase via constant mixing that drives rapid cooling of the recently mixed 'warm' gas (Fielding et al. 2020) and the hot phase. A constant hot wind blowing along the $\hat{y}$ (vertical) direction of the simulated domain gradually accelerates the cold phase (shown here in the monotonic growth of the velocity and displacement of the centre of mass). These results are consistent with the results presented in Banda-Barragán et al. (2021), for the high-res, short-box simulation, now supplemented with the results for the low-res, long-box simulation. We refer the reader to that paper for a comprehensive analysis of the evolution of the entire cold phase in the high-res model.

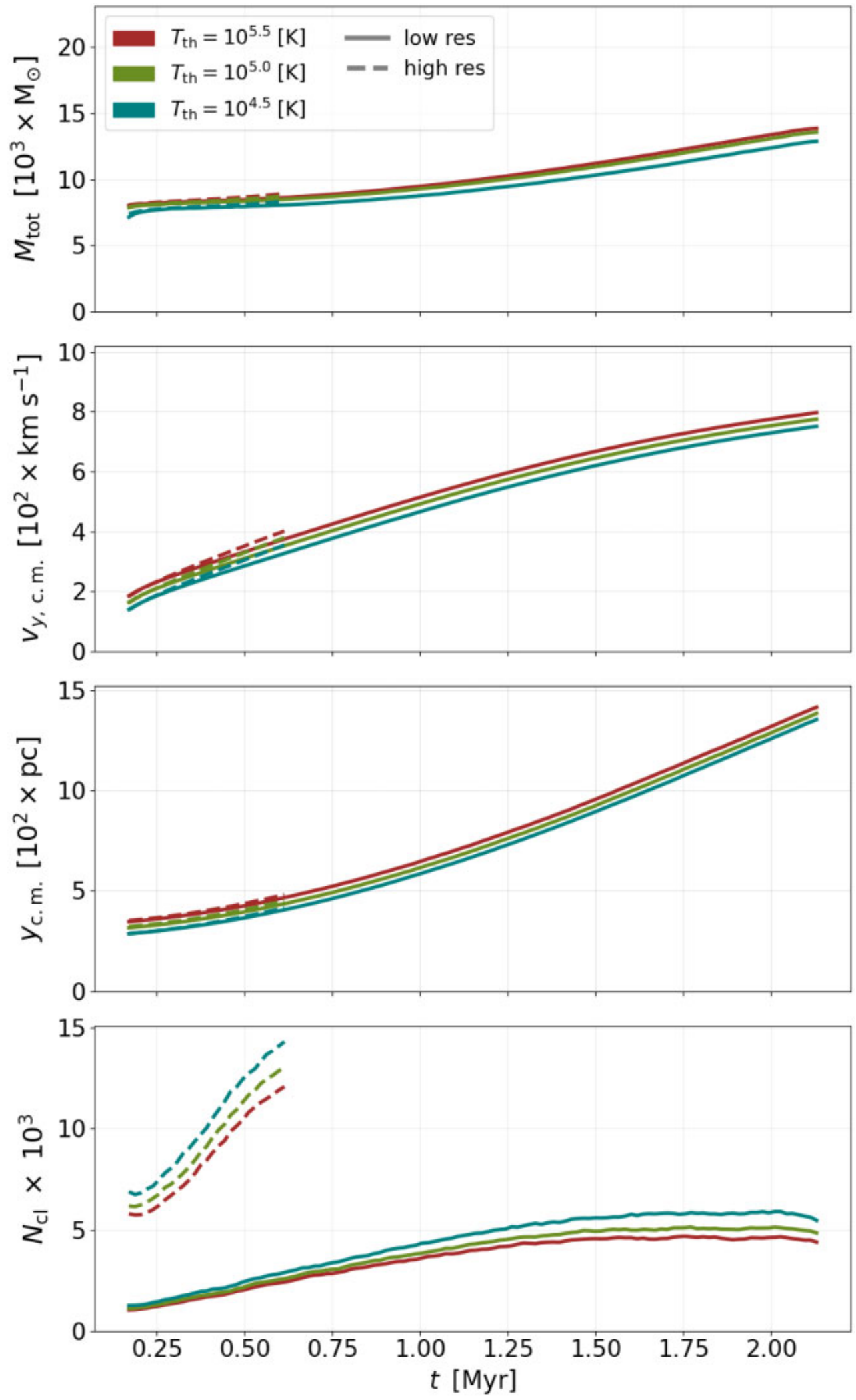

**Figure 4.** Evolution of the overall cold phase properties. Results are presented for the high-res simulation in dashed lines, and for the low-res in solid lines. For the highest temperature threshold, we tested $10^{5.5}$ K in red, intermediate $10^{5.0}$ K in green, and lowest temperature threshold $10^{4.5}$ K in blue. Top to bottom panels: total phase mass, cold phase centre-of-mass velocity in the direction of the shock $\hat{y}$, cold phase centre-of-mass position in the direction of the shock, and number of clouds received from the FoF analysis. The three curves in the velocity and location panels are practically identical, thus $10^{5.0}$ and $10^{5.5}$ K curves were shifted vertically for better visualization.

### 4.2 Finding clouds with our FoF algorithm

After identifying all the cells meeting the $T < T_{\rm th}$ condition, we employ our FoF procedure. In the rightmost panel of Fig. 1, we show the input binary map to the FoF routine at a particular snapshot. Darker regions within that panel meet the criterion $T < T_{\rm th}$ ($= 10^{5.5}$ K in this specific example). Our algorithm starts with a cell that meets our threshold criterion, and iteratively searches for neighbouring cells that fulfill the same threshold. This step is repeated until no more new neighbouring cells can be added, identifying a connected group. We refer to these groups as 'clouds' in our terminology, and sometimes, when we need to emphasize a gradation of size we also call them 'cloudlets' interchangeably. This procedure is repeated until all cells with $T < T_{\rm th}$ in the simulation are assigned to a specific group (cloud). The output from our FoF routine is a list of clouds per each analysed snapshot. Each cloud contains the list of cells that topologically belong to that cloud.

Fig. 3(a) shows a subsection of a 2D slice through the 3D temperature distribution, at $t = 1.5$ Myr, previously presented in Fig. 1. The iso-surface lines highlight two of the threshold values we examined in this paper ($10^{4.5}$ and $10^{5.5}$ K). Panel (a) in this figure shows the input into the FoF routine. The following panels (b)–(d) show 3D renderings of some examples of clouds detected by our FoF routine. The smallest cloud (panel b) has $\sim 6 \times 10^2$ cells; an intermediate cloud (panel c) has $\sim 6 \times 10^3$ cells, and the largest cloud (panel d) contains $\sim 8 \times 10^4$ cells. The 3D renderings were created using the 'PARAVIEW' 5.12.1 software.[2] Despite the 2-dex difference in size, these clouds share some morphological properties. They are all stretched along the $\hat{y}$ (wind) direction, and show a highly porous morphology with cold material very closely intertwined with hot gas. Inspecting renderings at other simulation times reveals that a self-similar, 'noodle'-like shape characterizes most clouds. We quantify such morphological properties in the next sections.

The output of our FoF routine can also be used to count the number of separate clouds ($N_{\rm cl}$; see the lower panel of Fig. 4), and their individual volumes (cell count per cloud). The output can further be used to tag individual clouds and use their coordinates to study other fields in the simulation, such as cloud averaged densities or velocities.

### 4.3 Number of clouds

The bottom panel of Fig. 4 shows the time evolution of the number of distinct clouds extracted via our FoF analysis. Initially, the entire multicloud layer is continuously connected. This occurs because most cells in the initial layer satisfy our temperature threshold criterion, $T_{\rm th}$. Subsequently, during shock propagation through the multicloud layer, the gas is strongly compressed by the shock and disrupted by dynamical instabilities. Large fractions of the cold material undergo 'destruction' through shear-driven mixing with the hot wind gas. This disruptive process, combined with fast recondensation (as shown by the thermodynamical analysis of the high-res simulation reported in Banda-Barragán et al. 2021, e.g. see their fig. 5), results in the break-up of the initial continuous layer into a few $\times 10^3$ separate unique clouds. Cooling times are very short, so warm gas can re-condense back efficiently. Before the shock traverses through the multicloud layer, the FoF analysis is dominated by transient effects. We therefore begin our analysis at $\sim 0.17$ Myr, a time by which the shock has entirely passed the initial layer. The FoF analysis is carried out at each snapshot up to $\sim 0.7$ Myr for the high-res, and up to $\sim 2.1$ Myr for the low-res simulation, times at which significant amounts of the material associated with the initial multicloud layer leave the computational domain.

The number of clouds, $N_{\rm cl}$, is of the order of a few $\times 10^3$ and monotonically grows with time (see Fig. 4, bottom panel).

[2] https://www.paraview.org/

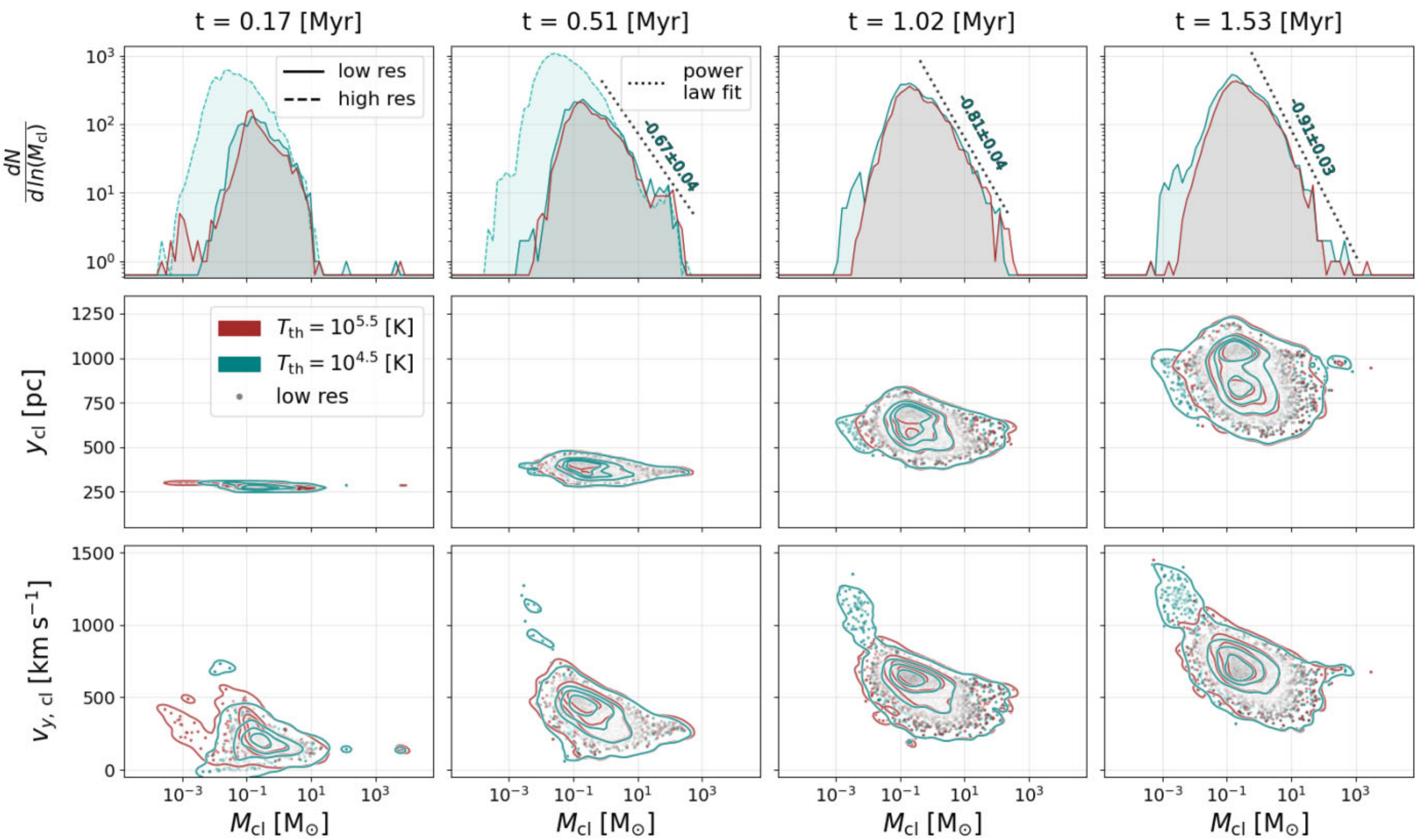

**Figure 5.** Results of the FoF analysis for two temperature thresholds. $T_{th} = 10^{5.5}$ K in red and $10^{4.5}$ K in blue, at a few selected snapshots in the evolution of the simulation. Top row of panels shows histograms of cloud number per mass range of clouds. Solid lines present the histograms of the low-res simulation, while the two leftmost panels also show for comparison, in dashed lines, the results of the high-res simulation. The top panels, for $t \geq 0.51$ [Myr], also show in dotted lines the best-fitting slope of a power law, fitted to the cloud number received for our low-res simulation, for $T_{th} = 10^{4.5}$ K and $M_{cl} > 2 \times 10^{-1}$ M$_\odot$. The best-fitting value, jointly with its $1\sigma$ uncertainty, is displayed above the dotted line, which has been slightly shifted to the right for better visualization. The central row shows contour lines representing the 0.5th, 25th, 50th, and 75th percentiles of the cloud positions along the shock direction ($\hat{y}$). The bottom row presents the velocity of the clouds in the $\hat{y}$-direction, using the same contour levels. The scatter plots represent individual clouds, with marker transparency inversely proportional to cloud density in that region of parameter space. For clarity, both the cloud position and velocity plots display results only from the low-res simulation.

As expected, the high-res simulation (dashed lines) hosts a higher number of clouds by a factor of $\sim 2^3$, than its low-res counterpart. This is due to the higher resolution of this simulation, which allows to resolve smaller scales, of the order of $\sim 1$ cell sized objects ($0.39^3$ pc$^3$ in high-res versus $0.78^3$ pc$^3$ in low-res). Notably, we also find that the number of clouds for the lowest temperature threshold is consistently higher than that for the highest $T_{th}$ across both resolutions by an order of $\sim 25 - 30$ per cent going from $10^{5.5}$ to $10^{4.5}$ K. This is due to connection of separate cold kernels via the linking length (similar to Sparre et al. 2019) as the temperature threshold is increased, which in turn reduces the number of discretely separable clouds. A visual demonstration of how the extension of the cloud boundaries, through an increase of the temperature threshold, can incorporate additional warm gas can be seen in the iso-surface lines presented in panel (a) of Fig. 3. There, some of the separate cold, $T < 10^{4.5}$ K, regions merge into a singular continuous region when looking at the higher threshold value of $T < 10^{5.5}$ K. Furthermore, this result unveils that a large fraction of cold clouds, defined as material falling under our lower threshold temperature $T < 10^{4.5}$ K, are not exposed directly to the post-shock hot flow ($\sim 10^7$ K). Instead, these cores are engulfed or 'cocooned' within warm gas shells, which fall under the higher threshold value we used $T < 10^{5.5}$ K, gas found at $\sim T_{mix}$ (similar to results reported by Nelson et al. 2020).

### 4.4 Cloud mass statistics: fragmentation, (re-)condensation, and coagulation

Fig. 5 presents 1D histograms of cloud counts (top panel), along with 2D histograms of cloud vertical displacements (middle panel), and cloud velocities (bottom panel) as functions of cloud mass. Contour levels in the 2D histograms correspond to the 0.5th, 25th, 50th, and 75th percentiles of the distributions. The cloud population, observed across all snapshots, spans over $\sim 5 - 7$ orders of magnitude in mass. Each column in this figure presents histograms at different times. The leftmost histograms show the cloud population at an early time, immediately after the shock has passed through the multicloud layer. At this time, the mass distribution of clouds is dominated by a single large cloud at $\sim 10^4$ M$_\odot$, which constitutes the majority of the cold phase. This represents the initial multicloud layer, which briefly stays spatially connected, with very little mass accounted in the smaller objects detached from it due to the recently passed shock. In the subsequent histograms (columns to the right), this large cloud disintegrates into smaller cloudlets under the influence of the hot post-shock wind. The dashed lines (top panels) show how a higher numerical resolution allows to resolve the cloudlets at the bottom of the scale, and explains the large difference seen in the total number of clouds (see Fig. 4). Despite this, the total mass of the entire cold phase is the same for both resolutions.

Fig. 5 supports the cloud connecting effect discussed above, when the threshold temperature is increased. The analysis of a snapshot for $T < 10^{4.5}$ K (in blue markers) shows a large number of cloudlets at lower masses, which 'disappear' once the same snapshot is analysed for the higher temperature $T < 10^{5.5}$ K (in red). The inclusion of the warm 'cocoon' gas in $T_{\rm th}$ reduces the number of cold kernels, as they become connected through the warmer intercloud material with other cold kernels, thus forming more massive singular clouds.

In the histograms in top panels of Fig. 5, it can be seen that the distribution of clouds number as a function of their mass grows and becomes wider over time. The masses range of $\sim 10^1 - 10^3\,{\rm M}_{\odot}$ that is almost void of clouds at the earliest snapshot, becomes continuously populated at later times. This smooth expansion of the distribution is the net result of several competing processes: (1) fragmentation of large clouds into smaller cloudlets, (2) condensation, and (3) coagulation.

Fragmentation of clouds is inherently a destructive process. Ram-pressure and dynamical instabilities cause sections of the cold cloud to mix with the hot wind, raising their temperatures to $T_{\rm mix}$ or $T_{\rm hot}$. In addition, shattering, driven by large pressure imbalances, can lead to the loss of sonic contact between different parcels of cooling gas (McCourt et al. 2018). These processes lead to the mixing and fragmentation of an initially large cloud into a spectrum of smaller cloudlets. The entire initial cold gas reservoir rapidly mixes with the background hot gas (see fig. 7 in Banda-Barragán et al. 2021), so the fragmentation process is dominant at early times in our simulations and re-organizes the morphology of existing dense gas. For instance, Fig. 5 shows a single large cloud of $\sim 10^4\,{\rm M}_{\odot}$ at 0.17 Myr, which has disintegrated in the subsequent panel by 0.51 Myr.

At later times ($\gtrsim 0.5$ Myr) in our simulations, condensation becomes the dominant process. Condensation occurs when warm-mixed gas or background hot gas lose thermal energy via radiative cooling to feed the cold phase. For the warm-mixed gas, this process involves the recondensation of gas that was originally in the multicloud layer, and became rapidly heated up, fragmented, mixed, and then cooled. Warm gas is the product of destruction of pre-existing cold gas via mixing with the hot wind. At $T_{\rm mix}$, the radiative cooling rate increases sharply, leading to the rapid recondensation of mixed gas back into the cold phase ($T < T_{\rm th}$). For the background hot gas, this process involves the cooling-driven accretion of hot gas (not originally in the multicloud layer) that becomes drawn and entrained into the cold flow. The effects of condensation are well demonstrated in the evolution of $M_{\rm tot}$ in Fig. 4, where the total cold gas mass steadily grows, eventually doubling from its initial value. Similarly, Banda-Barragán et al. 2021 (see their fig. 7, panel c) showed that some of the cold gas is directly accreted from the wind. This was demonstrated by excluding the passive scalar (which tracks gas originally in the multicloud layer) from their calculations of the dense-gas mass fractions.

Condensation processes can generate a continuous distribution of cloud masses. New, small, resolution-limited clouds continuously form from the warm-mixed gas and the hot background. These appear as a persistent cluster of the least massive objects in the velocity and position panels of Fig. 5. Additionally, pre-existing clouds can grow by accreting cold gas directly onto their surfaces through condensation. Over time, the combination of these two condensation channels can establish a continuous distribution of cloud masses.

Coagulation occurs when two or more pre-existing cloudlets merge within the multiphase flow, forming a single, continuous cold cloud. Unlike condensation processes, coagulation neither creates nor destroys cold gas; rather, it serves to restructure and reorganize the morphology of the cold phase. Coagulation can be driven by turbulent flows and radiative cooling (Gronke & Oh 2020). Pressure gradients and coalescence (Waters & Proga 2019) can lead to cloud mergers. When combined with condensation, coagulation can accelerate the formation of a smooth and continuous cloud mass distribution, as observed in the top panels of Fig. 5.

Two noticeable features in the middle and bottom panels of Fig. 5 are: (1) clouds with very distinct masses spatially co-exist, and (2) the cloud velocities $v_{\rm y,cl}$ are negatively correlated with mass. The more massive clouds, on average, move slower than the smaller ones. Although the positions of these very low-mass cloudlets remain centred within the general cloud population, their velocities appear unchanged over time and consistently exceed the typical cloud velocity. The reason for this is that the low-mass, high-velocity clouds observed at different snapshots do not represent the same cluster of clouds. A different population of low-mass clouds emerges at every time. This population is a reflection of the dynamic production of low-mass cold gas driven by the condensation of fast-moving warm and hot gas. These smaller and faster clouds, given sufficient time, are likely to be entrained by or coagulate to form larger clouds. We can define a typical time-scale for coagulation as $t_{\rm coag} \equiv \sigma_{\rm y}/\sigma_{v_{\rm y}}$, where $\sigma_{\rm y}$ stands for the standard deviation of the locations of the centres of mass of clouds, and $\sigma_{v_{\rm y}}$ is the standard deviation of the velocities of clouds in the wind direction. Using the results of our FoF analysis, we find that $t_{\rm coag}$ is proportional to the elapsed time of the simulation, $t_{\rm coag} \simeq t/2$. Such a model suggests that the coagulation rate may decrease over time due to the increasing spatial extent.

Overall, through the evolution of the simulation ($\gtrsim 0.5$ Myr), a steady-state population of clouds emerges. The mass distribution evolves into a persistent lognormal distribution, with a power law-like slope, which we discuss later in the text. The origin of this steady-state distribution is interesting. Our current analysis shows there is almost twice as much cold gas (in model low-res) at late times as there was at $t = 0$, which indicates efficient condensation. However, with our current numerical setup, we cannot fully separate some of the above processes and quantitatively weigh their individual contributions to the mass of dense gas, e.g. shattering versus other fragmentation mechanisms, cloud growth via gradual accretion or via coagulation, and different coagulation mechanisms. Integration of FoF analysis with distinct tracer particle routines in the simulation offers a potential avenue to disentangle these processes by establishing a precise cloud merger tree and its temporal evolution. By identifying specific clouds through snapshots according to their tracer particles, and by analysing the origin on newly accreted gas, one could potentially distinguish between coagulation of two clouds and gradual accretion directly from the hot component. In this suite of simulations, we did not employ a tracer particle routine, so we reserve this intriguing route of investigation to future work.

### 4.5 Cloud mass spectrum

The cloud mass spectrum (top panels of Fig. 5) deserves particular attention. The mass spectrum evolves in time, approaching a distribution akin to lognormal that peaks at $\sim 2 \times 10^{-1}\,{\rm M}_{\odot}$. For masses greater than the peak, particularly at late times, the shape of the spectrum resembles a power law, $dN/d\ln(M_{\rm cl}) \propto M_{\rm cl}^{a}$, with $a < 0$. Best-fitting values of this slope, for $T_{\rm th} = 10^{4.5}$, are shown as grey dotted lines, and are offset for clarity. The value of $a$ is printed above the fitted lines. As noted previously, the distribution starts from a narrow cluster of small clouds, and a singular massive cloud, at $t \sim 0.17$ Myr, immediately after shock passage through the initial multicloud layer. Later, the large cloud breaks into smaller clouds, while the smaller cloudlets grow in mass through accretion of fresh condensed gas and coagulation with other clouds. Overall a persistent continuous lognormal distribution emerges at $t \gtrsim 0.5$ Myr. Accordingly the slope of the power law evolves as well. Starting with $a \simeq -0.6$ at $t \sim 0.5$ Myr, dropping to a minimal (negative) value $a \simeq -0.9$ at $t \sim 1.2-1.5$ Myr, and slightly increasing to $a \simeq -0.8$ at later times ($t \sim 2$ Myr).

There is some debate in the literature regarding the mass spectrum of cold clouds. In an idealized suite of simulations of a single cloud in a wind tunnel, Gronke et al. (2022) reported a slope of $\sim -1$. At galaxy cluster scales, for a pure hydrodynamic simulation with AGN feedback, Li & Bryan (2014) found a slope of $\sim -3/4$, at the end of their simulation. Investigating clouds in wind tunnel simulations, Jennings et al. (2022) measured a large span of slopes, within the range $\in [-0.7, -0.84]$, once a steady-state population of clouds had formed. For zoom-in simulations of CGM regions of MW-like galaxies selected from TNG50 (Nelson et al. 2019; Pillepich et al. 2019), Ramesh & Nelson (2024) reported a slope of $\sim -1$. In addition, Tan & Fielding (2024) reported a similar slope in their galactic scale tall-box simulation setup where clouds emerge from the turbulent fragmentation of ISM disc gas as winds are launched.

All these simulations have differently sized computational domains, use distinct geometries and lengths for the simulation runs, and include/omit different physical processes. These differences are likely responsible for the different slopes of the cloud mass spectrum measured. Despite these, the reported power-law slopes are remarkably similar. To summarize broadly, in well developed and smooth distributions, the power-law slope ranges between the values of $\sim -2/3$ to $\sim -1$. Our simulations also span these values during the temporal evolution of our wind–multicloud flow. However, longer simulation times, better statistics, and an examination of varying initial conditions and of boundary conditions is still necessary to test whether these slopes are indeed universal.

## 5 MORPHOLOGY AND DYNAMICS OF CLOUDS

The results presented in Figs 4 and 5 reveal that a phase-scale examination (averaging over the entirety of the cold gas) of the velocities of the cold phase fails to capture the detailed motion of small clouds. They can travel $\sim 2-3$ times faster than the centre-of-mass velocity of the entire phase. Therefore, to better assess the entrainment of individual clouds, we wish to learn how their cross-sections evolve. Measuring the aerodynamicity of a given cloud, and its exact location within the density and velocity profile of the wind determines how well it is being accelerated by the wind (Klein et al. 1994), and through which physical process it acquires momentum. In addition, quantifying the surface area of a cloud with respect to its volume allows us to assess how elongated it is, to determine its fractal dimension, and to measure the dimensions of the mixing layer surrounding it (Fielding et al. 2020). The global morphology of the cold phase, i.e. whether it is spatially concentrated or spread out into a mist of small droplets (Gronke & Oh 2020), can be studied by calculating the surface area of individual clouds. Similarly, a measurement of the cross-section of these clouds exposed to the wind is essential for studying coagulation processes (Gronke & Oh 2023). In this section, we demonstrate how these questions can be addressed in a quantitative manner, using the results of our FoF analysis.

### 5.1 Surface area and fractal dimension

To study the surface area of individual clouds, we count the outward-facing facets of external cloud cells, and sort these surfaces based on their orientation. Fig. 6 presents the cloud surface areas as a function of cloud volume ($V_{\rm cl}$) at three distinct stages in the simulation. In the top row, markers with a darker colour represent the total surface area of the entire cloud ($S$), while contours and markers with a lighter colour indicate the cross-sectional surface area of the same cloud facing the wind ($A$). The latter is the combination of all the external surfaces of a cloud facing the $-\hat{y}$ direction. Similarly to our previous results, we observe a broad spectrum of cloud sizes.

Notably, the cloud surface and cross section grow along very narrow, power-law-like distributions on the $(S, V)$ and $(A, V)$ diagrams. The power-law relation can be expressed as: $S \propto V^{\alpha}$. In this case, $\alpha$ corresponds to one third of the fractal dimension $D$. For $l$, a typical length-scale of the cloud $S \propto l^D \propto V^{D/3}$ (Barenblatt & Monin 1983; Gronke et al. 2022). Typically, $\alpha$ carries the values of $\{5/6, 8/9\}$ as shown analytically in Mandelbrot (1975). In the top row of Fig. 6, the orange-shaded area represents the span of $\alpha$ values between 8/9 (top boundary of shaded area) and 5/6 (bottom). These fractal dimensions have previously been found in different simulation setups, e.g. 5/6 by Federrath et al. (2009), Fielding et al. (2020), Tan, Oh & Gronke (2023), and 8/9 by Federrath et al. (2009). Using the FoF results at each snapshot, we extract the value of $\alpha$ using a least-squares-method linear fit. Throughout the simulation, $\alpha$ has values between the two quoted fractions above, i.e. $\alpha \in [0.85, 0.875]$. The largest clouds in the population diverge to higher slopes of $\sim 0.9$. This behaviour is very similar to the one reported in Tan & Fielding (2024), with the largest clouds following a power law of $\sim 11/12$.

The top row of Fig. 6 also shows the expected surface area of a cloud (black dashed lines) and its cross-section exposed to the wind (black dotted line), given a spherical approximation for the shape of the cloud, $\propto V^{2/3}$. We find that the spherical approximation, $A_{\rm sph} = S_{\rm sph}/4\ (= \pi R^2)$, consistently overestimates the actual cross-sections for small- and intermediate-sized clouds. The clouds are aerodynamic with the cross-section facing the wind smaller than that implied from the spherical approximation. The spherical approximation also underestimates $A$ for the largest cloud(s) during early and late times (leftmost and rightmost panels), and approximately matches the exact value of $A$ at intermediate times (central panel). This varying degree of concordance with respect to the largest objects in the distribution is particularly noteworthy, given that the largest cloud(s) encompasses the majority of the cold phase mass and carries the majority of momentum. In Section 5.2, we use the precise cross-section measurements to evaluate the drag and momentum transfer between the hot and the cold phase.

The bottom row of panels in Fig. 6 illustrates the elongation of each cloud by depicting the ratio between its total surface and cross-section areas. The cloud surface area is a property highly dependent on numerical resolution, so it is typically regarded as an unresolved property in simulations (Fielding et al. 2020; Gronke & Oh 2020). Our findings support this view, as indicated by the substantial $S/A$ scatter observed for low-volume clouds. Conversely, as we transition to larger objects, the scatter in the ratio of surface areas diminishes. For the largest clouds in the system, the elongation converges to well defined values. The larger clouds within the population exhibit an elongation parameter of $S/A \sim 10$ at the initial stage of the evolution, reaching an order of $\sim 20$ by its end. Throughout this period, these values significantly surpass the classical spherical approximation of $S/A = 4$ (dashed lines), and seem to be inherent to the physical conditions we studied. Indeed, the better resolved clouds with larger volumes align themselves with the wind acquiring these characteristic elongation parameters.

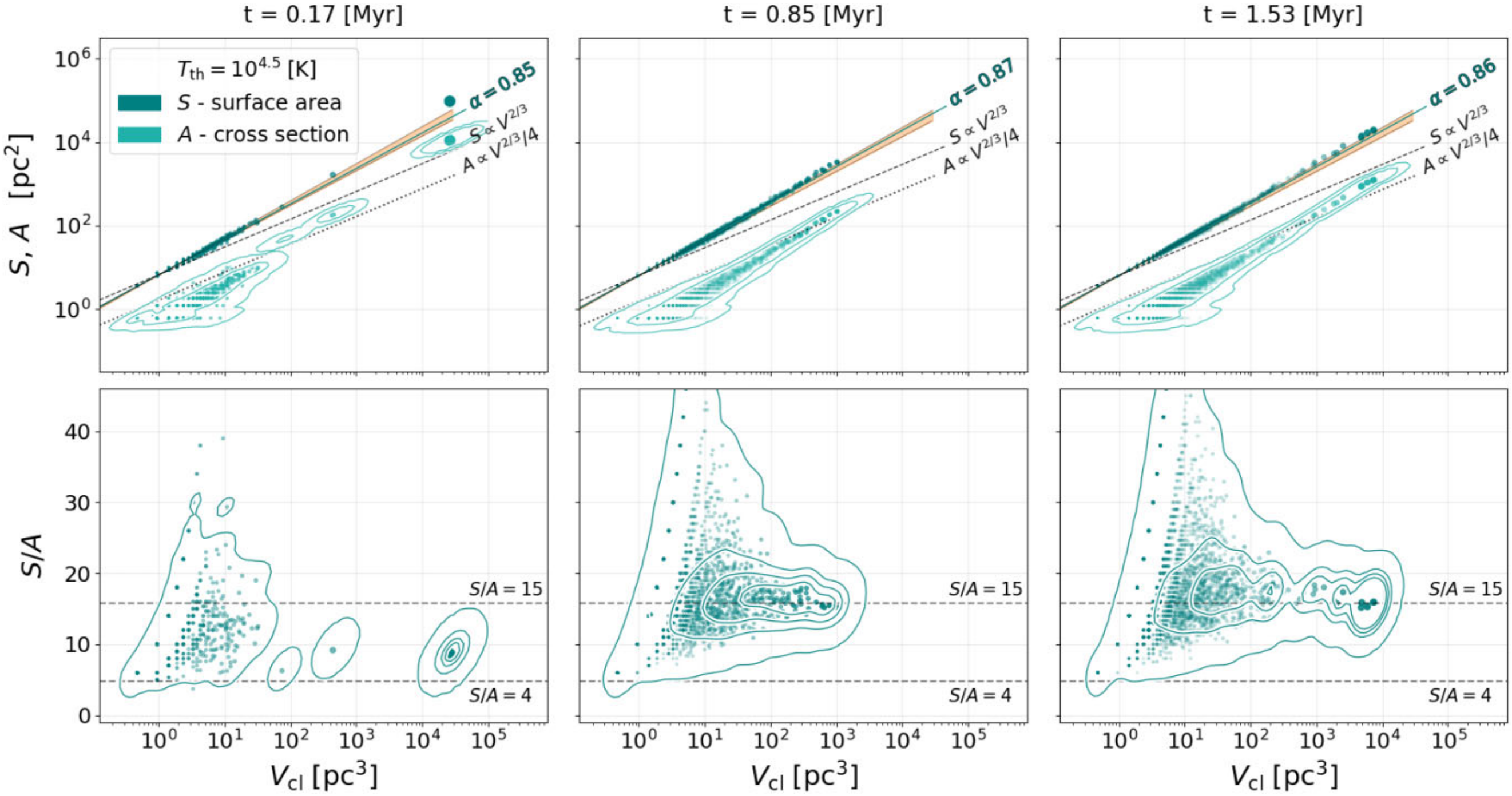

**Figure 6.** This figure presents the surface areas of clouds as identified using our FoF analysis in the low-res simulation, for a threshold temperature of $T_{\rm th} = 10^{4.5}$, K. Three times are shown: an initial time when the main shock front has just passed the initial multicloud layer (left), and two additional, more evolved times (centre and right). In the top row, the darker-shaded round markers represent the total cloud surface area $S$ as a function of cloud volume. This corresponds to the sum of all facets separating the 'cloud' from the 'wind' cells. The lighter-shaded contour lines indicate, for the same clouds, the 0.5th and 50th mass-weighted percentiles of the distribution of the sum of all cloud surface facets oriented in the direction opposite to the wind ($-\hat{y}$), denoted as $A$. Additionally, specific clouds are highlighted using round markers, with marker size proportional to cloud mass and transparency inversely proportional to cloud mass. The dashed black line in the top row represents the expected cloud surface, while the dotted line shows the wind cross-section, both assuming a spherical cloud shape. The blue–green line indicates the best-fit line, with the corresponding slope $\alpha$, while the orange-shaded region marks the range of slopes from a power law of 8/9 (top edge of the shaded area) to 5/6 (bottom edge).The bottom row presents the cloud elongation $S/A$, as a function of cloud volume, using the same presentation style as the top row. Contour lines indicate the mass-weighted 0.5th, 25th, 50th, and 75th percentiles of the distribution. The dashed horizontal lines mark both the classical spherical approximation $S/A = 4$, and the elongation parameter of $S/A = 15$, to which the system very visibly aligns, particularly for larger clouds.

### 5.2 Momentum and entrainment

In Fig. 4, we present the centre-of-mass velocity of the entire cold phase in the $\hat{y}$-direction as a function of time. Remarkably, for the three temperature thresholds and both resolutions, the phase-scale velocity remains practically identical. However, when we observe the velocities of individual clouds, a higher level of complexity in the system dynamics emerges. The bottom row in Fig. 5 shows the specific velocities of distinct clouds at particular times in the simulation. Apart from the clear signs of vertical motion, a substantial scatter in the velocities of the clouds exists. While the most massive few clouds dictate the large-scale kinematical behaviour, the smaller cloudlets can exhibit velocities that are a factor of $\sim 2-3$ higher than the centre of mass. A per-cloud analysis of acceleration can be very beneficial, as a phase-scale approach misses much of the complexity in the dynamics of clouds. In this section, we discuss the entrainment of the cold clouds in the hot wind. The results in Fig. 4 show that the cold phase in our simulation persists, monotonically grows with time, and is effectively accelerated along the $\hat{y}$-direction. Thus, we aim to answer the question of what the prevalent entrainment mechanism is, performing the analysis from a by-cloud perspective.

First, we define our terminology considering that the clouds in our wind–cloud systems have three phases. A cold phase $T_{\rm cold} < 10^{4.5}$ K, often spatially residing within a layer of warm-mixed gas $10^{4.5} < T_{\rm warm} < 10^{5.5}$ K, which in its turn is surrounded by the hot wind gas $T_{\rm hot} > 10^{5.5}$ K. Hot wind gas, which moves faster than the cold cloud and its mixing layer, interacts with the cocoon of warm gas around the cold phase (or with the cold cloud directly, without a warm intermediary), and accelerates it. Via this process, the cold cloud, and the warm gas that surrounds it become (partially or fully) comoving with the hot gas. We refer to this process as 'ram-pressure' acceleration. The acceleration process can enhance turbulence within the warm mixing layer, entraining more pristine hot gas into the mixing region. The temperature in the mixing region is $T_{\rm mix}$ ($= 10^{5.5}$ K, for: $T_{\rm cold} = 10^4$, $T_{\rm hot} = 10^7$K), at which the cooling rate rises sharply, effectively and quickly forming 'new', fast-moving cold gas via radiative cooling in the immediate vicinity of the cold cloud or directly on its surface. We refer to this additional process as 'condensation-driven' acceleration.

We attempt to differentiate between these two processes using a by-cloud analysis. Our prior assessment of the surface area $S$, and cross-sectional surface facing the wind $A$ of each cloud (see Section 5.1) is crucial for quantifying momentum transfer. In addition, we need an accurate measurement of the velocities and density profiles along the wind ($\hat{y}$) direction. As can be seen in Fig. 1, when hot post-shock gas encounters cold and slow clouds it moves along the path of lesser resistance and is funneled into channels created in between the cold clouds. Such channels are characterized by an increased velocity in the shock direction and low gas densities. Overall, this causes both the velocity and density of the wind to be non trivial and diverge from the constant post-shock values enforced at the domain boundary. We define 1D volume-weighted profiles for the velocity and density of the wind along the length of the simulated box as

$$v_{\mathrm{y,w}}(y) = \frac{1}{\mathcal{N}(y)} \sum^{\mathcal{N}(y)} v_y(x, y, z;\ T > T_{\mathrm{th}}), \tag{1}$$

$$\rho_{\mathrm{w}}(y) = \frac{1}{\mathcal{N}(y)} \sum^{\mathcal{N}(y)} \rho(x, y, z;\ T > T_{\mathrm{th}}). \tag{2}$$

Here, $\mathcal{N}(y)$ represents the number of cells at the $y$ coordinate that lie above $T_{\mathrm{th}}$. Owing to the well-defined wind direction (along $\hat{y}$), individual clouds develop negligible transverse velocities and momentum. When integrated over the entire cloud ensemble, these transverse components effectively vanish. For this reason, in this subsection, all velocity and momentum values are measured along the wind direction and are represented as scalars in the notation to highlight this behaviour.

The analysis is performed on the temporal derivative of the momentum of the cold phase, $\dot{p} = m\dot{v} + \dot{m}v$, with the left-hand side of the equation representing the total momentum change as is experienced and measured within the cold phase, and the right-hand side representing two evaluations of momentum change, accessible via our FoF analysis and measured from the perspective of the wind. The first partial derivative corresponding to momentum exchange due to velocity change, without mass transfer, we refer to as the ram-pressure term, $\dot{p}_{\mathrm{ram}}$. The second term, which includes transfer of mass into the cold phase without velocity change (comoving) we refer to as the condensation term, $\dot{p}_{\mathrm{cond}}$. Using the results of our FoF analysis, the amount of momentum lost by the wind per unit time via ram pressure is defined as

$$\dot{p}_{\mathrm{ram}} \equiv \left(\frac{\partial p_{\mathrm{w}}}{\partial t}\right)_{\mathrm{m}} = \sum_{i}^{N_{\mathrm{cl}}} \rho_{\mathrm{w}}(y_i)\, A_i\, u^2(y_i), \tag{3}$$

for the $i$th cloud out of $N_{\mathrm{cl}}$, with its centre of mass located at $y_i$, and a wind cross-section $A_i$. Here, the shear velocity between the cloud and the wind at the cloud's centre of mass is: $u(y_i) = v_{\mathrm{y,w}}(y_i) - v_{\mathrm{y},i}$, and the density of the wind is: $\rho_{\mathrm{w}}(y_i)$. We neglect an order of unity correction for the drag coefficient, $C_d$ that takes into account the fact that the area in the flow that is affected by the cloud is larger than its physical surface area, and by the fact that the hot wind does not fully stop and transfers all its momentum as it interacts with the cloud. Since the flow is subsonic with respect to the hot wind, we do not expect bow shocks to form, and the drag coefficient is expected to be below unity.

The second component of the momentum flux from the perspective of the wind is the momentum transferred via condensation from the warm to the cold phase. Ideally, by tracing the mass of individual clouds through time, one could calculate this mass accretion directly. Since our code is Eulerian, and we have not introduced tracer particles and constructed a cloud merger tree at this time, we resort to the following approximate method. We define the total condensation momentum transfer, taking into account specific cloud velocity, as

$$\dot{p}_{\mathrm{cond}} \equiv \left(\frac{\partial p_{\mathrm{w}}}{\partial t}\right)_{\mathrm{v}} = \sum_{i}^{N_{\mathrm{cl}}} \frac{S_i}{S_{\mathrm{tot}}}\, \dot{M}_{\mathrm{tot}}\, v_{\mathrm{y},i}, \tag{4}$$

for the cloud's centre-of-mass velocity $v_{\mathrm{y},i}$, with $\dot{M}_{\mathrm{tot}}$ representing the temporal change of the entire cold mass. The temporal derivative of the mass curve was computed after fitting the original curve to a high-order polynomial, so as to receive a smoother curve. Since mass accretion via condensation is a surface-dependent process, we chose to distribute the newly added mass between the cold clouds, weighing according to their respective surface with respect to the total surface of all the clouds in the population, $S_i/S_{\mathrm{tot}}$. We verified that distributing the mass to individual clouds according to their respective volumes and masses yields the same qualitative results. Finally, for the left-hand side of our momentum equation, we calculate the total change of momentum in the cold phase. This information is measured directly from the simulation, summing over the entire cold gas, and does not require a per-cloud analysis,

$$\dot{p}_{\mathrm{tot}} \equiv \frac{\mathrm{d}p_{\mathrm{tot,\,cl}}}{\mathrm{d}t}. \tag{5}$$

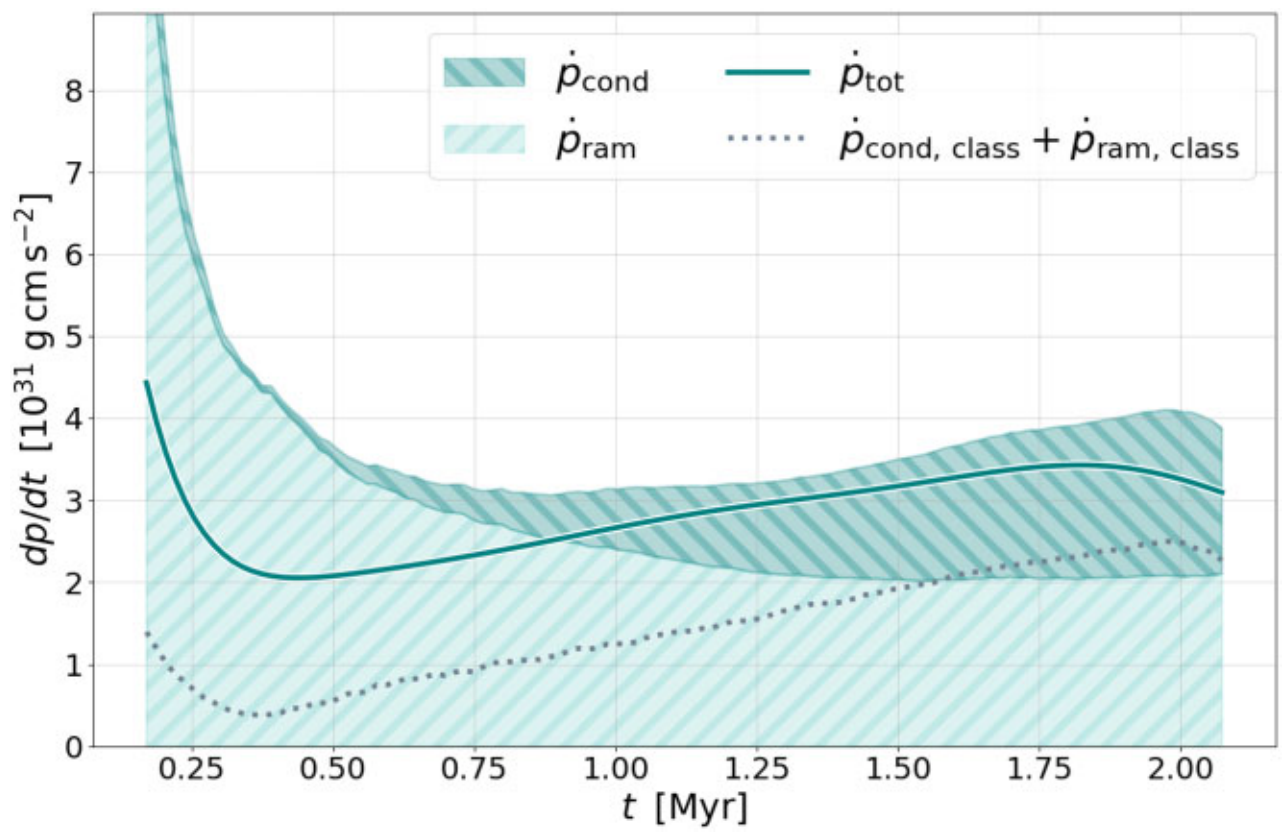

**Figure 7.** Temporal derivative of momentum as a function of simulation time. The solid blue line represents the total value of $\dot{p}$ as measured directly from the 'cold' gas cells with $T_{\mathrm{th}} = 10^{4.5}$ K. The dark dashed area shows the momentum change due to condensation of gas from the 'warm' to the 'cold' phase. The lighter dashed area shows the amount of momentum gained by the cold phase via ram pressure, using the exact cloud cross section extracted from FoF analysis. The dotted grey line shows an assessment of momentum change due to ram pressure, assuming a simplified spherical cloud geometry.

Our ram-pressure definition assumes that the entire momentum of the hot gas hitting the cloud is transferred to the cloud. More realistically, some of the hot gas gets diverted to flow around the cloud indicating a less than ideal efficiency, so our ram-pressure momentum transfer is an upper limit. We expect that the sum of ram-pressure and condensation momentum transfer to be greater or equal to the total momentum change of the cold component,

$$\dot{p}_{\mathrm{tot}} \leq \dot{p}_{\mathrm{ram}} + \dot{p}_{\mathrm{cond}}. \tag{6}$$

We present all the resulting momentum fluxes (equations 3, 4, and 5) in Fig. 7. The momentum contributions are shown in stacked, hatched areas: the lighter colour shows the contribution due to ram pressure, while the darker colour shows the one due to condensation. The evolution of the total momentum derivative is shown in a solid line. Note that at early times the momentum contribution to the cold phase via condensation is an order of magnitude smaller than one delivered via ram pressure. This aligns well with the initial evolution we observe in the spatial picture in the simulation (Fig. 1), where the strong shock passes through the initial multicloud layer. During this process (up to $\sim 0.17$ Myr) and for a short period of time after the shock has left the multicloud layer behind (up to $\sim 0.3$ Myr) the warm mixing layer material is being rapidly washed away and mixed into the hot wind. We see a reciprocal decrease in $M_{\mathrm{warm}}$ up to that point in time. At the same time, the cold clouds are practically stationary, thus experiencing maximal shear velocity at earlier times of the simulation. For these reasons, condensation has no warm mass to draw itself from, while the ram pressure is at its peak.

At later times, the mass of the warm gas gradually grows, and, with it, we can see an increase in the amount of momentum change associated with condensation of new comoving gas out of the warm phase. At the same time, cold clouds become entrained in the warm wind (though they do not reach full entrainment within the span of the simulation). The cross-section of clouds experiences a minor growth of up to a factor of $\sim 2$ (Section 5.1), and the ram pressure gradually decreases at later times. It is worth noting that at all times the momentum loss terms from the perspective of the wind exceeds the actual momentum change measured in the cold phase, $\dot{p}_{\rm ram} + \dot{p}_{\rm cond} > \dot{p}_{\rm tot}$. Failure to close this 'momentum budget' would have indicated that there is a principal error in this type of analysis. Out of the excess momentum, above what is actually measured in the cold phase, the drag coefficient can be evaluated, $C_{\rm D}$ (Klein et al. 1994). We measure it as: $C_{\rm D} = (\dot{p}_{\rm tot} - \dot{p}_{\rm cond})/\dot{p}_{\rm ram}$, indicating how much of the momentum available for entrainment actually made its way into the cold phase. In our simulation, $C_{\rm D}$ varies from its minimal value of $\sim 0.3$ at the initial shock-passage phase of the simulation, to its maximal value of $\sim 0.9$, at $t \sim 1.25$ Myr.

In Fig. 7, we also present (grey, dotted line) the 'classical' momentum change, based on equations (3) and (4), but for $\dot{p}_{\rm ram}$ we use a spherical approximation for the cloud cross-section and the centre-of-mass velocity of the entire cold phase (instead of individual clouds). Similarly, in the classical calculation of $\dot{p}_{\rm cond}$, we use the velocity of the centre of mass of the entire cold phase, and set the surfaces ratio to $S_i/S_{\rm tot} = 1$, as it is essentially a single-cloud calculation. The classical approximation qualitatively follows the $\dot{p}_{\rm tot}$ evolution, but it fails to 'close the momentum budget'. This approximation creates a substantial momentum deficit and fails to capture the dynamics of the system. At early stages, the deficit is $\dot{p}_{\rm tot}/(\dot{p}_{\rm cond,\,class} + \dot{p}_{\rm ram,\,class}) \gtrsim 4$, and at late stages, $\dot{p}_{\rm tot}/(\dot{p}_{\rm cond,\,class} + \dot{p}_{\rm ram,\,class}) \sim 2$.

To summarize, using our definitions of 'ram-pressure' and 'condensation-driven' accelerations to quantify the entrainment of the cold phase, we conclude that both contribute and that their relative importance depends on the stage of acceleration. At early times, the acceleration is clearly ram-pressure dominated, but as time progresses, condensation becomes at least equally important. We predict that at later times it could potentially outgrow the ram-pressure contribution.

## 6 THERMODYNAMIC STATE OF THE COLD PHASE

### 6.1 Thermodynamical time-scales

Our FoF analysis provides insights into the thermodynamic properties of distinct clouds. Computing typical time-scales helps us characterize the behaviour and evolution of the system. We focus on four such time-scales: three dynamical and one radiative. The dynamical time-scales are: (1) the sound-crossing time $t_{\rm sc} = l/c_{\rm s}$, where $l$ is the length of the cloud in its shortest dimension (usually transverse to shock), and $c_{\rm s}$ is the speed of sound in the cloud; (2) the cloud-crushing time $t_{\rm cc} \equiv \chi^{1/2} l/u$, which is a disruption time associated with the shock motion across the cloud; and (3) the growth time of KH instabilities, $t_{\rm KH} \sim t_{\rm cc}$, perhaps up to an order of unity correction. Additionally, the time-scale governing the creation of cold material through radiative cooling is: (4) the cooling time $t_{\rm cool} = 3k_{\rm B}T/2n\Lambda(T, n)$. Fig. 8 illustrates these typical time-scales relative to the cooling time, as a function of cloud's volume.

Evidently, for well-resolved clouds, $t_{\rm cool}$ stands out as the shortest time-scale in the problem by $\sim 1 - 5$ orders of magnitude. For the largest cloud(s) in the distribution the typical time-scales are of the order of: $t_{\rm cc} \simeq 10$, $t_{\rm sc} \simeq 5$, and $t_{\rm cool} \simeq 10^{-3}$ Myr. While the smallest, resolution limited clouds in the distribution have a substantial scatter, the typical values of their time-scales are: $t_{\rm cc} \simeq 10^{-1}$, $t_{\rm sc} \simeq 10^{-1}$, and $t_{\rm cool} \simeq 10^{-3}$ Myr. The transition of gas from the hot $> 10^6$ K to the cold phase $\lesssim 10^4$ K through radiative cooling is the fundamental process driving the creation of the cold phase. On the other hand, shock heating and dynamical instabilities are responsible for the reverse transition, i.e. for the destruction and mixing of cold gas. Since the cooling time is several dex shorter than the cloud-crushing and KH times, the dynamics of gas between the phases is predominantly dominated by the cooling of the warm, mixed gas phase via radiative processes.

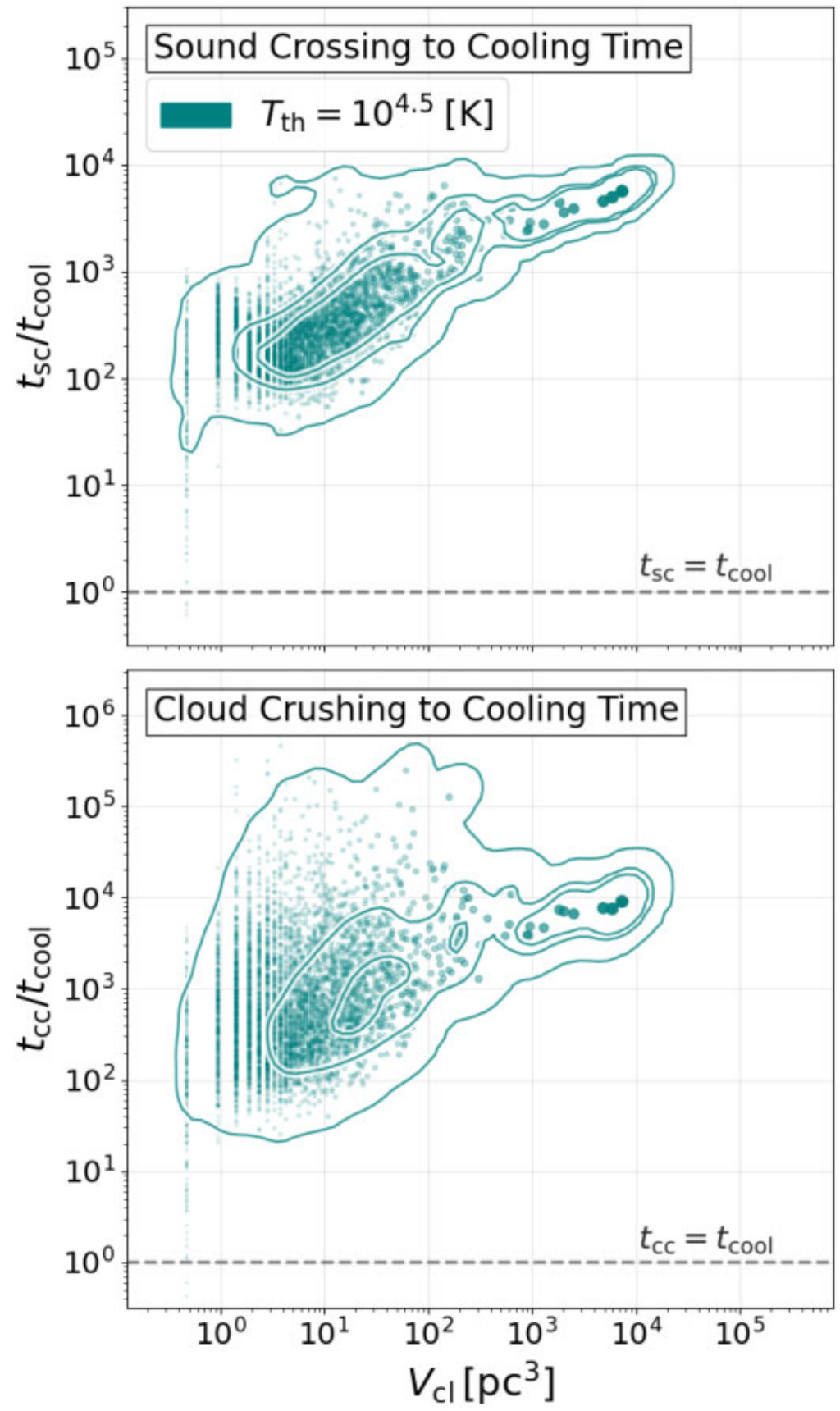

**Figure 8.** Typical time-scale ratios in our per-cloud analysis for a threshold temperature of $T_{\rm th} = 10^{4.5}$ K, evaluated at $t = 1.5$ Myr. Contours represent the 0.5th, 25th, and 50th mass-weighted percentiles of the cloud distribution. Round markers indicate individual clouds, with marker size proportional and transparency inversely proportional to cloud mass. The top panel shows the ratio of the cloud sound-crossing time to the cooling time as a function of cloud volume, while the bottom panel presents the ratio of the cloud-crushing time to the cooling time. Overall, the cooling time-scale is several dex. shorter than both dynamical time-scales, especially for larger and better-resolved clouds.

In large clouds, where the cooling time is shorter than the sonic crossing time by orders of magnitude, different areas of the cloud are out of causal contact with the majority of the cloud. The resulting large ratio of $t_{\rm sc}/t_{\rm cool} \gg 1$ would support a rapid fragmentation and shattering of clouds. However, the results depicted in Fig. 5 show a contrasting scenario. Instead of witnessing fragmentation, larger clouds persist throughout the simulation. These clouds exhibit mass growth, forming a 'rain'-like structure along the post-shock flow. This behaviour supports condensation and coagulation, which together lead to a continuous increase in the number of large clouds and their mass. The persistent brevity of the cooling time across all resolved clouds throughout the simulation suggests that the cold material within them does not align with the expected cooling–heating balance curve.

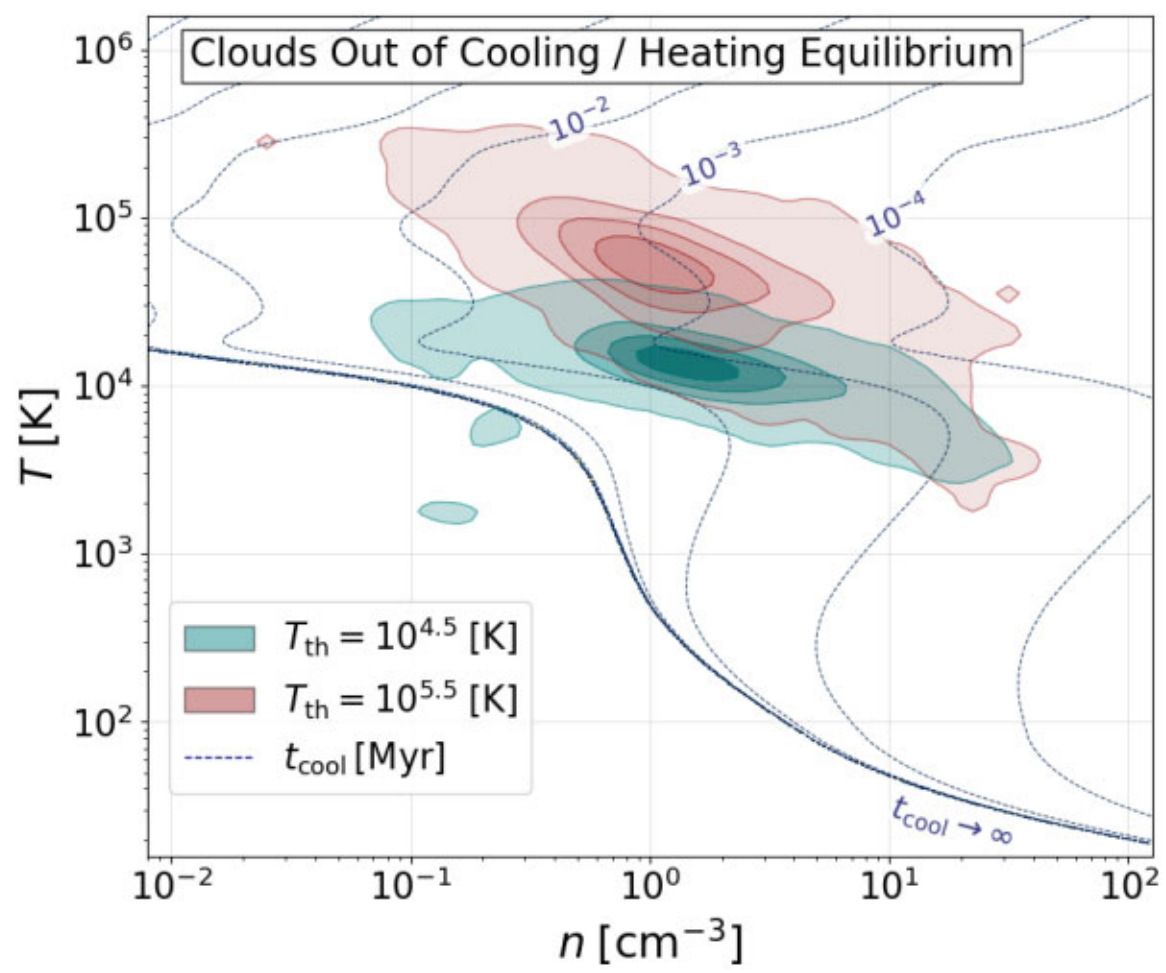

**Figure 9.** 2D histogram showing the distribution of clouds as a function of mass-weighted mean temperature and number density of each cloud, for two threshold temperature values $T_{th} = 10^{4.5}$ and $10^{5.5}$ K, at $t = 1.5$ Myr. Contour levels represent the 5th, 25th, 50th, and 75th percentiles of each distribution. The dark-blue dashed lines indicate several values of the cooling time-scale $t_{cool}(T, n)$ in Myr. Evidently, the cold clouds remain stable for extended periods of time $\sim$ Myr, in areas of the phase space where the cooling time is notably very short $\sim 10^{-3}$ Myr.

Fig. 9 supports this observation by presenting dashed lines that indicate cooling time values $t_{cool}$ in Myr, as a function of temperature and density. Additionally, the figure shows the distribution of clouds for two threshold temperature values, plotted as a function of mean cloud temperature and gas number density. As the net cooling time increases, the cooling time lines approach the heating–cooling equilibrium state, $|\Lambda(T, n) - \Gamma(T, n)|$ (Fig. 2). Below this region, the gas experiences more efficient heating from the background UV radiation than radiative cooling, rendering the cooling time-scale irrelevant (omitted for clarity). The minima in $|\Lambda(T, n) - \Gamma(T, n)|$ acts as a natural attractor in the system. Throughout the entire simulation (i.e. for $t \sim 2$ Myr), the majority of the clouds remain well above the expected balance line of our cooling–heating curve. This occurs despite the very short cooling times, $t_{cool}/t \sim 10^{-2} - 10^{-3}$, at which the vast majority of clouds are positioned on the $(T, n)$ diagram. The cold phase is held at a new thermodynamically stable state at a higher temperature than the one dictated by the cooling–heating function. As we discuss next, this constant loss of energy through radiative cooling is compensated by another source (dissipation of internal cloud motions).

### 6.2 The role of dynamic pressure and shocks

To understand the role of internal cloud motions in mediating the thermal balance in our wind–multicloud systems, we look into the residual velocities within our FoF-detected clouds. The velocity of each cloud cell, relative to its centre-of-mass motion $\boldsymbol{v}_{c.m.,cl}$, is defined as the local residual velocity

$$\boldsymbol{v}_{res}(x, y, z) \equiv \boldsymbol{v}(x, y, z) - \boldsymbol{v}_{c.m.,cl}. \tag{7}$$

It is important to note that $\boldsymbol{v}_{res}$ contains within itself a contribution from five different components of motion within each cloud: (1) the velocity of turbulent motion that evolves naturally within a cloud. (2) Residual velocities carried over from a merger of several smaller cloudlets into a single cloud. (3) Motions that originate from the edges of the cloud, where it interacts with the hot wind, caused by both compression, as well as due to the pressure gradients created in the vicinity of cooling gas. (4) Large-scale rotational motion of the cloud. (5) Changes in cloud geometry, such as shortening or elongation of the cloud in the wind direction. We further explore and quantify the relative importance of the different velocity components within $\boldsymbol{v}_{res}$, and compare the significance of our centre-of-mass definition of $\boldsymbol{v}_{res}$, to the more commonly used definition of $\boldsymbol{v}_{turb}$ based on convolution, in Appendix A.

Regardless of their exact nature, these random internal velocities contribute to the effective dynamic pressure of the cloud ($P_{dyn}$), which supports it against the ambient pressure. Once the local residual velocity is known, the dynamic pressure in the cloud can be assessed as

$$P_{dyn} \equiv V_{cl}^{-1} \sum_{i \in cl} \boldsymbol{v}_{res,i}^2 \, \rho_i V_i \, / \, 2, \tag{8}$$

summing over all cloud cells. Here, the density of a cloud cell is $\rho_i$, the cell volume is $V_i$, and cloud's total volume is $V_{cl}$. Similarly, the thermal pressure of a cloud is defined as

$$P_{th} = V_{cl}^{-1} \sum_{i \in cl} n_i \, k_B \, T_i \, V_i, \tag{9}$$

and the dynamic Mach number is computed as a volume-weighted root-mean-square (RMS) of the residual local velocity over the local sound speed,

$$\mathcal{M}_{dyn} \equiv \sqrt{V_{cl}^{-1} \sum_{i \in cl} (\boldsymbol{v}_{res,i}/c_{s,i})^2 \, V_i}. \tag{10}$$

The left panel of Fig. 10 illustrates the ratio between the dynamic and thermal pressure within the clouds. For the larger clouds (50th and 75th percentile lines) this ratio is of the order of $\sim 10$. This indicates that the primary pressure support in such clouds is their internal motion, with minimal thermal support. This result aligns well with prior studies, which also relied on per-cloud analyses. For instance, Nelson et al. (2020) showed that clouds were thermally underpressured in their CGM simulations, with the majority of the support delivered by the magnetic counterpart. In Tan & Fielding (2024), the clouds were found to be supported equally by thermal and turbulent pressure. Similarly analytic CGM profiles constructed to match the observed ionization and thermodynamic state of the gas by Faerman & Werk (2023) also require a considerable fraction of the pressure (3–10 times the thermal pressure) within the cold clouds to be non-thermal. This result implies that the internal gas motion within the clouds in our models is supersonic relative to the local gas. The central panel displays the typical $\mathcal{M}_{dyn}$ per cloud, confirming this result. This panel reveals that the majority of clouds are consistently supersonic, with typical $\mathcal{M}_{dyn} > 1$, and even values of $\mathcal{M}_{dyn} \gtrsim 10$ for the larger and well-resolved clouds. We note similar values reported by Warren et al. (2024), for the larger clouds formed in the CGOLS galactic tall-box suite of simulations (Schneider & Robertson 2018), with turbulent Mach numbers of $\mathcal{M}_{turb} \sim 2 - 10$. Please note that in Fig. 10 we compare the mean dynamic pressure within a cloud divided by the mean thermal pressure, as opposed to the mean of local pressures ratio. The latter would simply show the square of the dynamic Mach number, $\langle P_{dyn}/P_{th} \rangle \propto \mathcal{M}_{dyn}^2$.

Supersonic internal cloud motions also drive shocks, which are effective at transferring energy from the kinetic to the thermal component. Thus, we searched for shock waves using our own PYTHON-based, shock-finding routine (see Armijos-Abendaño et al. 2020; Navarrete, Pinargote & Banda-Barragán 2024). First, our algorithm searches for converging flows, i.e. for cells where $\nabla \cdot \boldsymbol{v} < 0$, and for large pressure gradients, i.e. for cells where $\nabla P/P > \epsilon$, where $\epsilon$ is chosen to filter out subsonic waves. This step returns a list of shocked cell candidates. Next, we use the Rankine–Hugoniot jump conditions to calculate the shock Mach numbers, $\mathcal{M}_{shock}$, in each shocked cell by considering the local velocity jumps (i.e. the differences between post- and pre-shock velocities, see Vazza et al. 2011) and the pre-shock sound speeds. We obtain the local velocity jumps using directional speed gradients (see equation 13 in Banda-Barragán et al. 2020 for strong shocks and equation 1 in Navarrete et al. 2024 for general shocks). This step returns 3D arrays containing all the detected shocks with their corresponding Mach numbers.

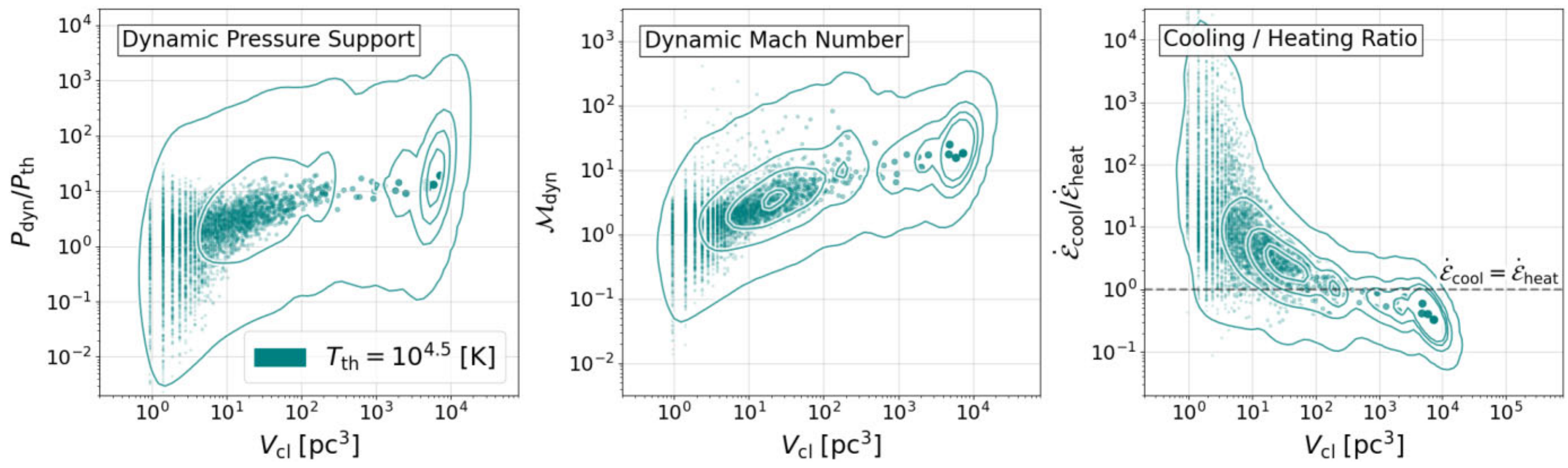

**Figure 10.** Cloud thermodynamics as unveiled by our FoF analysis for a threshold temperature of $T_{th} = 10^{4.5}$ K, at $t = 1.5$ Myr. Contours indicate the 1st, 25th, 50th, and 75th percentiles of the mass-weighted cloud distribution, while round markers denote individual clouds, with marker size proportional and transparency inversely proportional to cloud mass. The left panel shows the ratio of dynamic to thermal pressure as a function of cloud volume. The central panel presents the dynamic Mach number, while the right panel displays the ratio of energy lost via cooling to the energy available for heating through dynamic pressure dissipation. Overall, dynamic pressure, as opposed to thermal pressure, supports the cold clouds in the multiphase flow.

The results of our shock-finding routine, at $t = 1.5$ Myr, are shown in Fig. 11. The top panel shows a 2D histogram of the volume fraction of shocked cells as a function of the shock Mach number and the temperature of each cell. A large number of cells is found in a highly shocked regime. Shock Mach numbers of $\mathcal{M}_{shock} \simeq 2 - 10$ characterize gas at $T \sim 10^4$ K, and even higher shock Mach numbers of $\mathcal{M}_{shock} \sim 50$ are observed for the very cold gas at $T \sim 10^2$ K. In addition, the 'ridge' of the distribution of $\mathcal{M}_{shock}$ follows a negative correlation with respect to the gas temperature of $\propto T^{-1/2}$. This shows that the distribution of residual velocities within the clouds remains roughly constant at $\sim 25 - 30\,\mathrm{km\,s^{-1}}$ (which correspond to cloud internal FWHM, full width at half-maximum velocities, presented in Section 7 and Fig. 12), with the temperature of the gas setting the high $\mathcal{M}_{shock}$ values. The bottom panel of Fig. 11 shows the ratio of shocked cells to the total number of gas cells per each temperature bin, $f_{\mathcal{M}_{shock}}(T) \equiv \sum V_i(T; \mathcal{M}_{shock} > 1)/\sum V_i(T)$, where $V_i$ is the cell volume. We find that the dominant majority, $> 70 - 80$ per cent, of cold and warm gas, $T \lesssim 10^{5.5}$, is shocked. This further confirms that (supersonic) internal motions and the strong shocks associated with them are indeed a widely present phenomenon. Shear supersonic motions and shocks induce strong gas compressions, which raise local densities in the mixed phase and allow for recondensation to operate and reform cold gas. As a result, such cold, dense gas inherits some of the dynamic properties of the mixed phase, and the observed long-lived, fast-moving, multicloud flow arises.

We note that recent studies on phase transitions in the ISM report similar Mach numbers for turbulent gas clouds in the cold

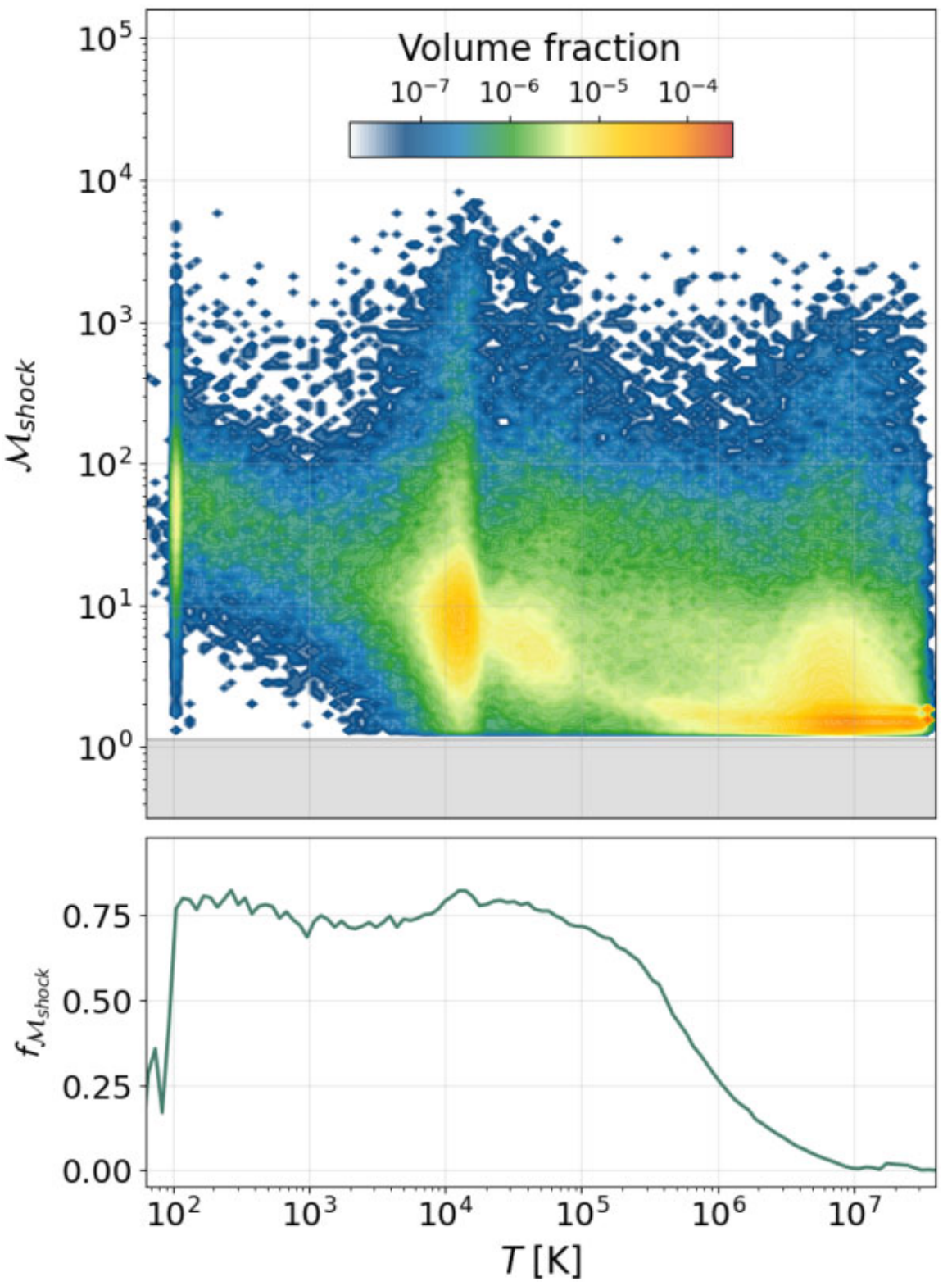

**Figure 11.** Shock distributions in our wind–multicloud models. The top panel presents a 2D histogram of all gas in the simulation at time $t = 1.5$ Myr, showing the local shock Mach numbers calculated by our shock-finding algorithm as a function of gas temperature. The shock-finding routine finds only cells with $\mathcal{M}_{shock} > 1$, so no subsonic values are shown in the histogram. The bottom panel shows the ratio of shocked cells to the total amount of cells at a given temperature.

(CNM) and warm neutral (WNM) media. Gerrard et al. (2024) finds turbulent Mach Numbers between $\sim 0.5 - 2$ in H I clouds with WNM temperatures $< 8000$ K, which are embedded in the nuclear wind of our Galaxy. Our analysis suggests a broader range of turbulent Mach numbers (with higher values also possible) for both the CNM and WNM phases, but overall our Mach number ranges engulf those reported by the observational study.

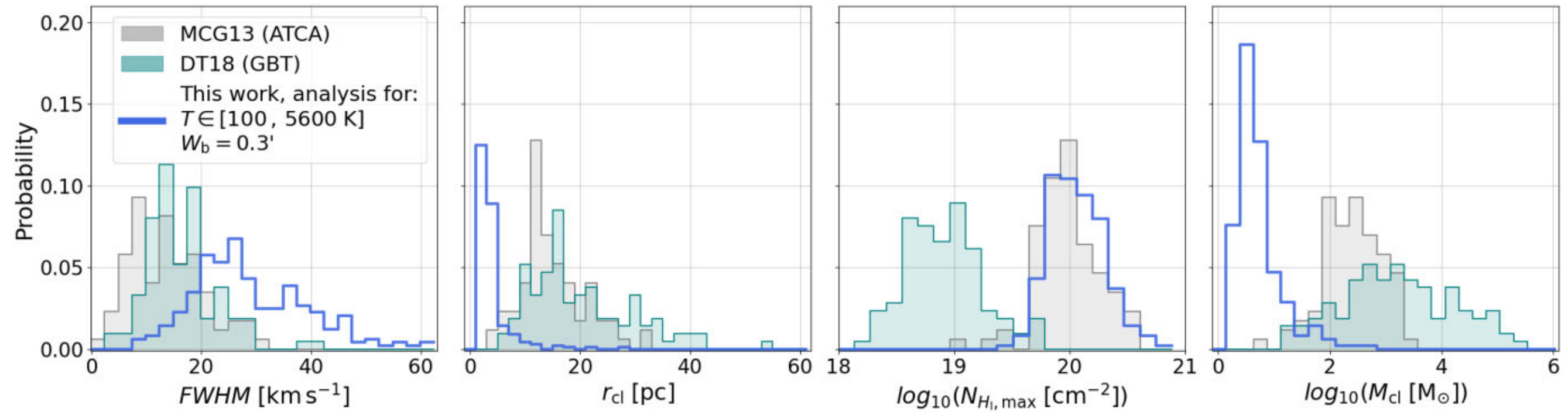

**Figure 12.** Histograms of the observed properties of the H I clouds detected in McClure-Griffiths et al. (2013) in grey, and Di Teodoro et al. (2018) in green. The solid blue line shows the H I clouds detected in our low-res simulation at $t = 2$ Myr. For clarity of presentation each histogram was normalized by its respective total cloud number. From left to right, we show the FWHM velocities within each cloud, the typical cloud radius, the maximal observed H I column density, and the cloud mass. Overall, there is a large discrepancy between the three data sets; two observations and one numerical, in the observed clouds' radii, column densities, and mass.

In order to quantitatively assess the energetic balance within the cold clouds we estimate how much energy is radiated away via cooling, and how much energy is dissipated by the shocks driven by the residual motion in cloud's frame of reference. The calculation of the energy flux lost by a cloud due to radiative cooling can be straightforwardly derived from the cooling function,

$$\dot{\mathcal{E}}_{\rm cool,cl} = \sum_{i\in {\rm cl}} n_i^2 V_i \, \Lambda(n_i, T_i). \tag{11}$$

Except for a small cluster of single-cell sized cloudlets, which are located in the background radiation heating regime (see Fig. 9), the majority of cold clouds are found in the radiative cooling regime of the $|\Lambda - \Gamma|$ function. For this reason, the $\Gamma(T, n)$ part is omitted in the definition of $\dot{\mathcal{E}}_{\rm cool,cl}$. The amount of kinetic energy available to produce heating can be assessed using the local residual velocity as

$$\dot{\mathcal{E}}_{\rm heat,cl} = d^{-1} \sum_{i\in {\rm cl}} |\boldsymbol{v}|^3_{{\rm res},i} \, \rho_i V_i, \tag{12}$$

where $d$ is the length of a single cell in the simulation. The ratio between the energy fluxes, $\dot{\mathcal{E}}_{\rm cool}/\dot{\mathcal{E}}_{\rm heat}$ is depicted in the right panel of Fig. 10 (Faerman et al. 2025 used a similar type of analysis to constrain the eddy size of turbulent cold gas in the CGM). Due to numerical resolution limitations, little or no residual motion can be resolved for smaller clouds, causing $\dot{\mathcal{E}}_{\rm cool}/\dot{\mathcal{E}}_{\rm heat} \gg 1$. As we transition to larger and better-resolved clouds, this ratio decreases sharply to $\dot{\mathcal{E}}_{\rm cool}/\dot{\mathcal{E}}_{\rm heat} \lesssim 1$. These clouds shape the overall phase-scale behaviour as they are the largest objects in the phase and carry the majority of the mass.

To sum up, the entire cold phase remains in equilibrium at temperatures higher than those defined solely by the radiative cooling–heating function. These conditions require an equilibrium: $\dot{\mathcal{E}}_{\rm heat} \sim \dot{\mathcal{E}}_{\rm cool}$. The hot wind continuously drives supersonic motion (for large clouds $\mathcal{M}_{\rm dyn} \gtrsim 10$) into the cold clouds, which contributes to $\dot{\mathcal{E}}_{\rm heat}$. There is continuous heating resulting from internal supersonic motions and their subsequent dissipation within the clouds. Additional simulations with other wind Mach numbers are required to demonstrate the generality of these results.

## 7 OBSERVATIONAL SIGNIFICANCE

### 7.1 Two H I cloud populations in the MW nuclear wind?

To complement our theoretical analysis, we discuss the relevance of our numerical work for observations. A per-cloud FoF analysis of simulations can be more naturally compared to observed clouds than a general phase-space analysis. The computational setup of our simulations was motivated by the gas conditions in the inner region of our Galactic CGM. In this section, we compare the results of our FoF analysis to observations of 'warm' H I clouds in the MW's nuclear wind. We focus on two observational studies by McClure-Griffiths et al. (2013, McG13) and Di Teodoro et al. (2018, DiT18), who report two populations of H I clouds at distances of up to 1.5 kpc above and below the Galactic plane. McG13 report 86 distinct clouds with LSR velocities of $\lesssim 200$ km s$^{-1}$, and DiT18 report 106 clouds with LSR velocities $\lesssim 300$ km s$^{-1}$. The temperature of these clouds is estimated to be $T \lesssim 4000$ K in McG13, and $T \lesssim 5600$ K in DiT18.

An interesting observational result is that the two H I cloud samples appear to have distinct estimated (maximal) column densities $N_{\rm H_I,max}$, which raises the question of whether they belong to different populations, distinct evolutionary stages of the nuclear wind, separate feedback events, or to the same population. In the latter case, can the differences be attributed to the spatial resolution and sensitivity of the radio telescopes used to detect them? In principle, both surveys, McG13 and DiT18, are expected to be observing clouds with similar properties, created and driven by similar physical processes. Di Teodoro et al. (2018) address this discrepancy (see their section 3.4), suggesting that the order of magnitude difference in the column densities and order of a few difference in cloud radii originate from the different spatial resolution and sensitivity between the two radio telescopes. In this case, the masses are expected to be very different as well, being the product of observed column density and cloud surface.

To answer this question, we resort to our FoF routine and carry out a systematic comparison of four metrics in both observations and the low-res simulation for gas with temperatures $T[{\rm K}] \in [100, 5600]$, these metrics are: $\Delta v$, $M_{\rm cl}$, $r_{\rm cl}$, and $N_{\rm H_I,max}$. Fig. 12 shows histograms of cloud counts normalized by the size of each respective data set for each of these metrics. The colour schemes and bins were taken to be similar to fig. 5 in Di Teodoro et al. (2018). Superimposed on top of the histograms, we present the result of our FoF analysis for H I gas. The clouds were taken from a late stage in the simulation at $t \sim 2$ Myr, so that the clouds had sufficient time to grow in mass and numbers. First, we compare the cloud velocity dispersions (leftmost panel of Fig. 12). Both surveys report similar values of $\Delta v \sim 15\,\mathrm{km\,s^{-1}}$, while the simulation produces clouds with higher internal velocities $\Delta v \sim 25\,\mathrm{km\,s^{-1}}$. The difference is likely due to the higher spectral resolution of our simulation compared to those of the observations, which are evaluated by measuring the width of the spectral emission line. Next, we compare the cloud radii (middle-left panel). The cloud radii reported in DiT18 were on average larger and reached higher maximal values than in McG13. The cloud radii in the simulation on average are much smaller $\sim 5$ pc versus $\sim 15$ pc in McG13. The difference here can be attributed to the higher spatial resolution of the low-res simulation, compared to those of the H I observations. Additionally, our simulations do not include magnetic fields, which in reality could contribute to larger cloud radii.

The H I column densities (middle-right panel) measured in our simulation and McG13 agree very well, with a slight tendency to higher values in the simulation results. On the other hand, the column densities measured in DiT18 are lower on average by 1 dex. We also compare the cloud masses (rightmost panel). The cloud masses presented in the two observations vary quite substantially $M_{\mathrm{cl,McG13}}/\mathrm{M}_{\odot} \in [10^{0.7}, 10^{3.5}]$ and $M_{\mathrm{cl,DiT18}}/\mathrm{M}_{\odot} \in [10^{1.1}, 10^{5.5}]$, with the results of the simulation producing on average much smaller masses, with only a few clouds reaching the highest span of masses of McG13, and at least two orders of magnitude lower than the maximal masses span of DiT18. As we show below, the differences in column densities and radii in the two cloud populations can be attributed to the radio instruments used to detect them. Indeed, McG13 reports observations carried out with the Australia Telescope Compact Array (ATCA) observatory, while DiT18 used the Green Bank Telescope (GBT). There is a substantial difference in the angular resolution of both telescopes, e.g. their effective beam widths are $W_{\mathrm{b,ATCA}} = 2.4$ arcmin versus $W_{\mathrm{b,GBT}} = 9.5$ arcmin. In addition, their minimal column density sensitivity thresholds are $10^{19}\,\mathrm{cm^{-2}}$ for ATCA and $1.5 \times 10^{18}\,\mathrm{cm^{-2}}$ for GBT.

### 7.2 A single H I cloud population: FoF reconciliation

We set out to check if the differences in the telescope sensitivities and spatial resolutions can bridge the gaps between the data sets. First, we project the 3D density onto the $(x,y)$ plane of the simulation, creating 2D column densities $N_{\mathrm{H_I}}$. After the projections of all FoF cloud pixels onto a 2D surface, the surface of all the pixels $A_{\mathrm{2D}}$ was calculated. From the surface of this mock observation, we then extracted the typical cloud radius $r_{\mathrm{cl}} = \sqrt{A_{\mathrm{2D}}/\pi}$, as was done in McG13 and DiT18. The reported distance to the Galactic centre of 8500 pc for McG13 (similarly 8200 pc for DiT18), for a cell size in the low-res simulation of $d_{\mathrm{cell}} \simeq 0.78$ pc, gives the simulation an effective beam width of $W_{\mathrm{b,sim}} \simeq 0.3$ arcmin, compared to $W_{\mathrm{b,ATCA}} = 2.4$ arcmin and $W_{\mathrm{b,GBT}} = 9.5$ arcmin. In the next step, the $N_{\mathrm{H_I}}$ distribution was convolved with a square top-hat function, $7 \times 7$ pixels large to mimic the spatial resolution of ATCA ($26 \times 26$ pixels for GBT). In order to avoid overlapping clouds and counting multiple clouds as a single object in the projection, we only projected a single cloud at a time. For the background value, we used the column density of the hot wind in the simulation $N_{\mathrm{H_I,wind}} = 10^{18}\,\mathrm{cm^{-2}}$. We also tested the results for different background column densities, varying all the way to zero, and the results were found to be insensitive to this value. The observed surface of a cloud $A_{\mathrm{2D}}$ was taken to be all the pixels in which the column density after convolution was above the minimal $N_{\mathrm{H_I}}$ detection threshold, $10^{19}\,\mathrm{cm^{-2}}$ for ATCA ($1.5 \times 10^{18}\,\mathrm{cm^{-2}}$ for GBT). Finally, the clouds, which at the end of this process, were found to be with a surface area $A_{\mathrm{2D}}$ smaller than the corresponding beam width were filtered out as well. The typical radius of a cloud was then calculated assuming a circular geometry from the adjusted surface. Subsequently, the maximal column density was taken as the maximal value found within the pixels of $A_{\mathrm{2D}}$.

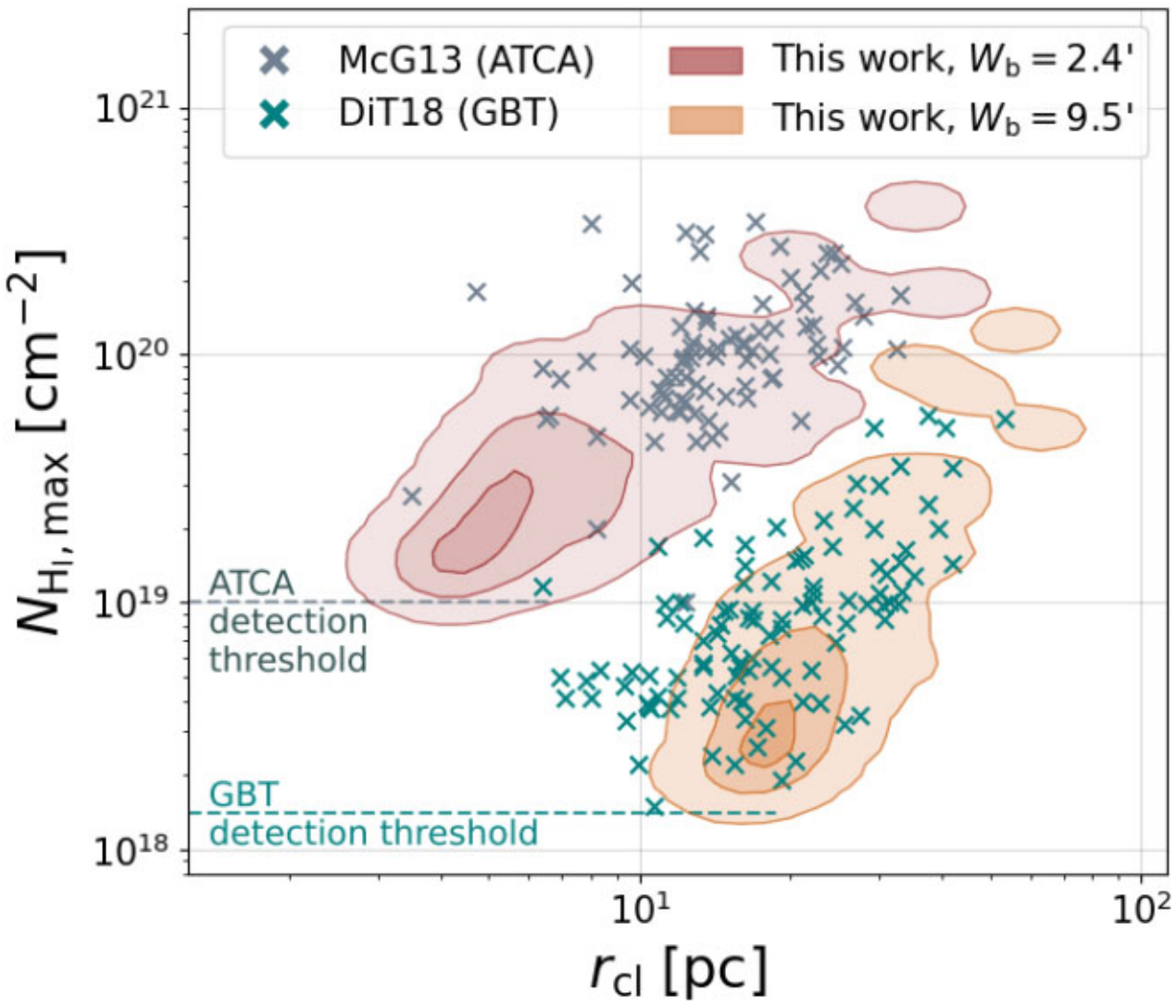

**Figure 13.** Correlation plot between the maximal observed H I column density and the typical radius of the clouds reported in McClure-Griffiths et al. (2013), grey x's, and Di Teodoro et al. (2018), green x's. The minimal column density detection threshold per each observation is shown in dashed lines of corresponding colours. The red and orange contour lines represent the 1st, 25th, and 75th percentiles of the cloud distribution from our FoF analysis of column densities and radii. These results have been adjusted to effective beam widths of $W_b = 2.4$ arcmin (comparable to McG13) and $W_b = 9.5$ arcmin (comparable to DiT18). The results correspond to the low-res simulation at $t = 2$ Myr.

Fig. 13 presents the results reported in McG13 and DiT18 in a correlation plot (in green and grey x's). The maximal observed column density $N_{\mathrm{H_I,max}}$ is displayed as a function of a typical observed cloud size $r_{\mathrm{cl}}$. In the same figure, we present contour plots showing the 1st, 25th, and 75th percentiles of our mock observation results, with two values of resolution reduction to the effective beam width of $W_b = 2.4$ arcmin to mimic ATCA, and $W_b = 9.5$ arcmin for GBT.

Fig. 13 presents the results of the low-res simulation at a later evolutionary stage of $\sim 2$ Myr. While both the observational data and the two resolution-corrected simulation data sets follow a similar overall trend, they are not perfectly cocentred. At the selected time, the overlap of the simulated data with the GBT data is more significant than with the ATCA data. For this time, our simulations predict smaller clouds with lower column densities compared to the ATCA data. Despite this, our analysis with telescope sensitivities and spatial resolutions does close the gap between the two sets of observations. Moreover, as discussed in Section 4.4 and 4.5, the lower end of the cloud volume distribution is limited by resolution constraints, whereas larger clouds continue to form over time through condensation and coagulation. Thus, at later times, beyond the scope of this simulation, we expect the distribution of simulated clouds to shift towards larger volumes/radii, and exhibit higher maximum column densities. This evolution would likely improve the overlap on the $N_{\mathrm{H_I,max}}/r_{\mathrm{cl}}$ plane.

Overall, the analysis of our simulations suggests that the effects of distinct telescope sensitivities and spatial resolutions can reconcile the differences in column densities and cloud radii between the two sets of H I observations. At the same time, the overlap indicates that the observed cloud population may trace a well-evolved steady-state population ('a boundary condition problem') and is independent (or not as dependent) on the specific wind launch conditions and the initial condition at the ISM (that would imply 'an initial condition problem'). Cloud–wind simulations provide an excellent laboratory to study the effect of initial conditions on the evolution and observational signature of the cold phase. Another important result is that the cloud radius–column density correlation, which essentially probes the density, is fitted naturally in the simulations and indicates that a considerable part of the pressure support in observed clouds comes from non-thermal pressure components. This non-thermal pressure support is also required to fit absorption line column densities in the UV (Faerman & Werk 2023).

## 8 CONCLUSIONS

In this paper, we studied the morphological, kinematic, and thermodynamic properties of dense clouds in galactic winds using a set of wind–multicloud simulations, analysed jointly with our novel analysis method that identifies specific clouds using a rolling-threshold FoF algorithm. Our FoF routine, written in PYTHON, searches for topologically connected clouds within the complex multiphase gas that result from the interaction between a hot wind and a layer of cold gas with a lognormal density distribution. After the initial stage of cloud identification, we perform additional analysis on individual clouds to study their interaction with the wind and other clouds, and to report a wealth of cloud-specific physical insights. Our main findings are summarized below:

(i) *On the cold phase structure of wind–multicloud flows.* Our FoF analysis reveals that cold clouds with $T \lesssim 10^4$ K are compact, but are not directly embedded in the hot wind. Instead, cold clouds are 'cocooned' within intermediate warm gas (i.e. a mixing layer) with $10^{4.5} < T < 10^{5.5}$ K, which shields them from direct interaction with the hot wind (consistent with results by Farber & Gronke 2021). Counting the number of clouds as a function of time $N_{\rm cl}(t)$ for different threshold temperatures $T_{\rm th}$ (Fig. 4) indicates that $N_{\rm cl}(t)$ decreases with increasing $T_{\rm th}$, which implies that warm, mixed gas connects the separate cold clouds via the 'linking length' (similar to results by Sparre et al. 2019). Similarly, our cloud population mass histograms (Fig. 5) show that low $T_{\rm th}$ produces a population of low-mass clouds, which 'disappears' for higher $T_{\rm th}$. We find that $\sim 25$ per cent of the clouds are linked with others via the warm gas 'cocoon'.

(ii) *On the morphology and fractality of cold clouds.* Cold clouds exhibit a self-similar morphology with a universal, 'noodle-like', elongated shape emerging at $\sim 100$ cells (Fig. 3). Large clouds are the spatial amalgamation of increasingly smaller cloudlets, all with a basic elongation parameter of $S/A \simeq 15$. The fact that this seems to be a very well defined property of the large and well resolved clouds suggests that $S/A$ is a characteristic property of wind–multicloud systems. The dependence of $S/A$ on the boundary conditions (e.g. shear velocity, density contrast, and cold phase initial geometry) will be studied extensively in future work. Additionally, our FoF analysis shows that the cloud surface area $S$ and cross-sectional area $A$ follow power-law-like distributions (see Fig. 6). For the surface area, we find $S \propto V^{\alpha}$ with $\alpha \in \{5/6, 8/9\}$, which have also been reported in previous studies by Mandelbrot (1975), Federrath et al. (2009), Fielding et al. (2020), Tan et al. (2023), and Tan & Fielding (2024) for distinct simulation setups.

(iii) *On cloud coagulation.* Our analysis shows evidence of substantial coagulation between cold clouds as the simulation evolves. This process is pivotal for the gradual mass growth of cold gas, which progressively and continuously populates the entire mass spectrum between the smallest (resolution-limited) cloudlets and the largest, most massive (resolved) clouds. We find that the coagulation time-scale $t_{\rm coag} \equiv \sigma_y/\sigma_{v_{\hat{y}}}$ is proportional to the simulation time $t_{\rm coag} = at$ for $a \sim 0.5$. In order to explore coagulation more explicitly and possibly even follow the mergers or breakage of specific clouds between consecutive snapshots, we suggest adding numbered tracer particles in individual clouds in future simulations. This would allow FoF analyses to tag and track the evolution of clouds in different snapshots.

(iv) *On momentum transfer and the cold phase entrainment.* Our FoF routine provides precise measurements of the cross-sections of clouds and the location of their interaction surface with the hot wind. These allow us to study the cold phase kinematics and entrainment in unprecedented per-cloud detail. Our analysis reveals that the transfer efficiency of momentum from the hot to the cold phase, as quantified by the drag coefficient, ranges from $C_{\rm D} \simeq 0.3$ to 0.9. Secondly, the simplified spherical model for clouds gives an adequate qualitative description of the momentum change of the system, but fails to close its momentum budget. The accurate measurement of wind-to-cloud interaction surfaces captures much better the momentum transfer process and manages to account for the entirety of the actual measured momentum change of the cold phase (Fig. 7). Lastly, we find that the dominant regime of momentum transfer between the hot wind and the cold clouds changes in time. Ram-pressure acceleration is the dominant process during the early stages of evolution, and condensation-driven momentum transfer becomes increasingly more dominant as time progresses. We reiterate the fact that our simulation does not reach a state of a fully comoving cold phase with the wind, and only captures the evolution partially. The dominant regime of momentum transfer should be studied more extensively in longer simulations in additional physical and geometrical setups.

(v) *On dynamical pressure and cold gas thermodynamics.* Having accurate measurements of the volume and geometry of each cold cloud allowed us to accurately calculate dynamical time-scales. Our FoF analysis confirms that the cooling times of cold clouds are shorter by a few dex than any other dynamical time-scale, such as $t_{\rm sc}$, $t_{\rm KH}$, $t_{\rm cc}$, and the time of the simulation $t$. Cold clouds are expected to cool rapidly (aided by strong compression) and shatter to a mist-like morphology, with a typical cloud size of $\sim c_s t_{\rm cool}$. Despite this, the vast majority of the clouds 'hover' above the saddle point in our heating–cooling function throughout the entire span of the simulation. We find that the dissipation of kinetic energy injected by the hot wind into the warm and cold phases via shocks and supersonic shear motions can balance highly efficient cooling. Supersonic internal motions within the clouds with $\mathcal{M}_{\rm dyn} \gtrsim 1$ for smaller resolved clouds, and values of up to $\gtrsim 10$ for the larger ones, together with a population of strong shocks are prevalent in the cold phase. The conversion of kinetic energy from the residual motion within a cloud's frame of reference through shocks into heat is an effective mechanism to offset the energy loss due to fast radiative cooling. For large and well-resolved clouds, we find that the amount of energy dissipated per unit time from these motions is a factor of a few larger than the thermal energy lost by a cloud due to radiative cooling (Fig. 10). A new thermodynamically stable state, at higher temperatures than the ones dictated purely by radiative processes, is maintained in the cold phase. Motivated by the steepness of the cooling function around $T \geq 10^4$ K, we predict that this behaviour is general and that the non-thermal to thermal pressure ratio will not depend strongly on the shear velocity. However, this point must be verified in future work.

(vi) *On the properties of $H_I$ clouds in the MW wind.* Using our FoF routine to study H I-emitting gas with $T$[K] $\in$ [100, 5600], we demonstrate that two samples of H I clouds observed in the nuclear wind of our Galaxy (McClure-Griffiths et al. 2013; Di Teodoro et al. 2018) very likely belong to the same population. By implementing a resolution reduction routine to match the effective beam width and the minimal column density sensitivity threshold of the ATCA and GBT radio telescopes, we produce mock H I observations. We find that the substantial differences in H I cloud radii and column densities between the two observed samples can be attributed to instrumental differences rather than physical ones. After the spatial resolution and sensitivity corrections, our simulated data reconcile both data sets on the $N_{\mathrm{H_I,\,max}}/r_{\mathrm{cl}}$ diagram (Fig. 13).

This paper illustrates the power of combining multiphase CGM gas simulations with cloud-finding FoF post-processing algorithms. Carrying out a phase-scale analysis reveals the overall behaviour of the cold phase, but does not provide details on the multiscale structure of wind–multicloud flows. On the other hand, a FoF analysis provides valuable insights into the fractal nature of cold gas. Applying this technique to numerical simulations with different initial conditions (e.g. other wind Mach numbers and density distributions), additional physics (e.g. models with electron thermal conduction and magnetic fields), and at other spatial scales (e.g. small-scale single wind–cloud and large-scale disc–wind models) would be essential to better comprehend the complexity of galactic winds and the generality of our results.

## ACKNOWLEDGEMENTS

The authors thank the anonymous referee for their insightful comments and suggestions, which helped significantly improve the clarity and quality of this work. We also thank Chen Zorea and Yakov Faerman for their valuable discussions and comments, which greatly contributed to improving this study.

The authors gratefully acknowledge the Gauss Centre for Supercomputing e.V. (www.gauss-centre.eu) for funding this project by providing computing time (via grant pn34qu) on the GCS Supercomputer SuperMUC-NG at the Leibniz Supercomputing Centre (www.lrz.de). This research was supported by the Munich Institute for Astro-, Particle and BioPhysics (MIAPbP) which is funded by the Deutsche Forschungsgemeinschaft (DFG, German Research Foundation) under Germany's Excellence Strategy – EXC-2094–390783311. We also thank the developers of the PLUTO code and the PYFC library, which we used to run our simulations. WEBB is supported by the National Secretariat of Higher Education, Science, Technology, and Innovation of Ecuador, SENESCYT. YB and OG were supported by Israel Science Foundation (ISF) grant 2190/20. MB acknowledges funding by the Deutsche Forschungsgemeinschaft (DFG, German Research Foundation) under Germany's Excellence Strategy – EXC 2121 'Quantum Universe' – 390833306. CF acknowledges funding provided by the Australian Research Council (discovery projects DP230102280 and DP250101526), and the Australia-Germany Joint Research Cooperation Scheme (UA-DAAD). We further acknowledge high-performance computing resources provided by the Leibniz Rechenzentrum and the Gauss Centre for Supercomputing (grants pr32lo, pr48pi, and GCS Large-scale project 10391), the Australian National Computational Infrastructure (grant ek9), and the Pawsey Supercomputing Centre (project pawsey0810) in the framework of the National Computational Merit Allocation Scheme and the ANU Merit Allocation Scheme.

## DATA AVAILABILITY

The data and the analysis code underlying this article will be shared on reasonable request to the corresponding author.

## APPENDIX: COMPARISON BETWEEN TURBULENT AND DYNAMIC MACH NUMBERS

In Section 6, we define the dynamic properties of the clouds. Summing over all cells of a cloud, at $\boldsymbol{r} = (x, y, z)$, the dynamic Mach number of a cloud is defined as:

$$\mathcal{M}_{\rm dyn} \equiv \sqrt{(V_{\rm cl})^{-1} \sum_{i\in {\rm cl}} \left(\boldsymbol{v}_i(\boldsymbol{r}) - \boldsymbol{v}_{\rm c.m.,cl}\right)^2 c_{{\rm s},i}^{-2}\, V_i}, \tag{A1}$$

with capital $V_i$, $V_{\rm cl}$ for volumes, not to be confused with the lower case $v$ which are velocities.

$\mathcal{M}_{\rm dyn}$ captures the residual motions within a cloud with respect to its centre-of-mass velocity. It comprises several components of the motion, all of which we wish to include in our analysis. First are the turbulent velocities within a cloud. These are caused by naturally occurring turbulence in a high Reynolds number flow, as well as driven compressionally by the strong interaction between the hot wind 'channels' that are created between the cold and slow(er) moving clouds. The second contributor is the constant merging of smaller cloudlets into larger clouds. For a typical temperature of $\sim 10^4$ K, the local speed of sound is $c_s \sim 10\,{\rm km\,s^{-1}}$, while the difference of velocities between clouds can reach an order of $\sim 100\,{\rm km\,s^{-1}}$ (Fig. 5). For this reason, we expect that a large portion (even a majority) of cloudlet mergers drive supersonic motion into the newly formed merged cloud. The third segment of residual velocities within a cloud originate from regions located at the edges of the cloud which experience substantial contraction due to interaction with the hot wind outside, as well as the influence of pressure gradients caused by the fast cooling of gas at the cloud-adjacent turbulent mixing layers. The fourth contribution to the motion can be accredited to large-scale rotation of the clouds. The fifth and final component is the large-scale geometrical changes of the cloud, such as stretching of the cloud in the wind direction, which we do observe, and wish to include in our calculation.

All these are cloud-internal motions and contribute to the kinetic energy budget of each cloud, that, given supersonic motions (that are indeed observed: see Figs 10 and 11), convert kinetic energy to thermal energy through shocks, and maintain the cloud at a heightened temperature with respect to the minimal temperature dictated purely by the heating–cooling function.

In order to quantify the relative importance of each of these velocities to the measured value, $\mathcal{M}_{\rm dyn}$, we compare the total to the turbulence-only component, which we estimate in the 'standard' way (Abruzzo, Fielding & Bryan 2024) of a density weighted (Favre averaging) filtering of small-scale kinetic power by real-space convolution:

$$\boldsymbol{v}_{i,{\rm turb}}(\boldsymbol{r}) \equiv \boldsymbol{v}_i(\boldsymbol{r}) - \langle \boldsymbol{v}_i(\boldsymbol{r})\rangle = \boldsymbol{v}_i(\boldsymbol{r}) - \frac{\iiint f_\sigma(\boldsymbol{r}-\boldsymbol{r}')\rho(\boldsymbol{r}')\boldsymbol{v}_i(\boldsymbol{r}')\,{\rm d}^3\boldsymbol{r}'}{\iiint f_\sigma(\boldsymbol{r}-\boldsymbol{r}')\rho(\boldsymbol{r}')\,{\rm d}^3\boldsymbol{r}'} \tag{A2}$$

A density-weighted convolution is performed with a Gaussian kernel $f_\sigma(\boldsymbol{r})$ with width $\sigma$. Then, the convolved velocities are subtracted from the original velocities field, to receive only the local residual velocities. This method is expected to leave only the local velocity fluctuations, such as turbulence and compression, while removing large-scale motions, such as bulk motion, stretching of the cloud or its rotation. As the cloud gas is $\sim 10^2 - 10^4$ times denser than the hot wind material, the density averaging strongly favours cells that belong to the cold clouds, and allow for minimal confusion between the hot gas into the cold phase. The turbulent Mach number is calculated using the filtered turbulent velocity, combined with our FoF cloud identification:

$$\mathcal{M}_{turb} \equiv \sqrt{(V_{cl})^{-1} \sum_{i \in cl} \boldsymbol{v}_{turb}^2 \, c_{s,i}^{-2} \, V_i}, \tag{A3}$$

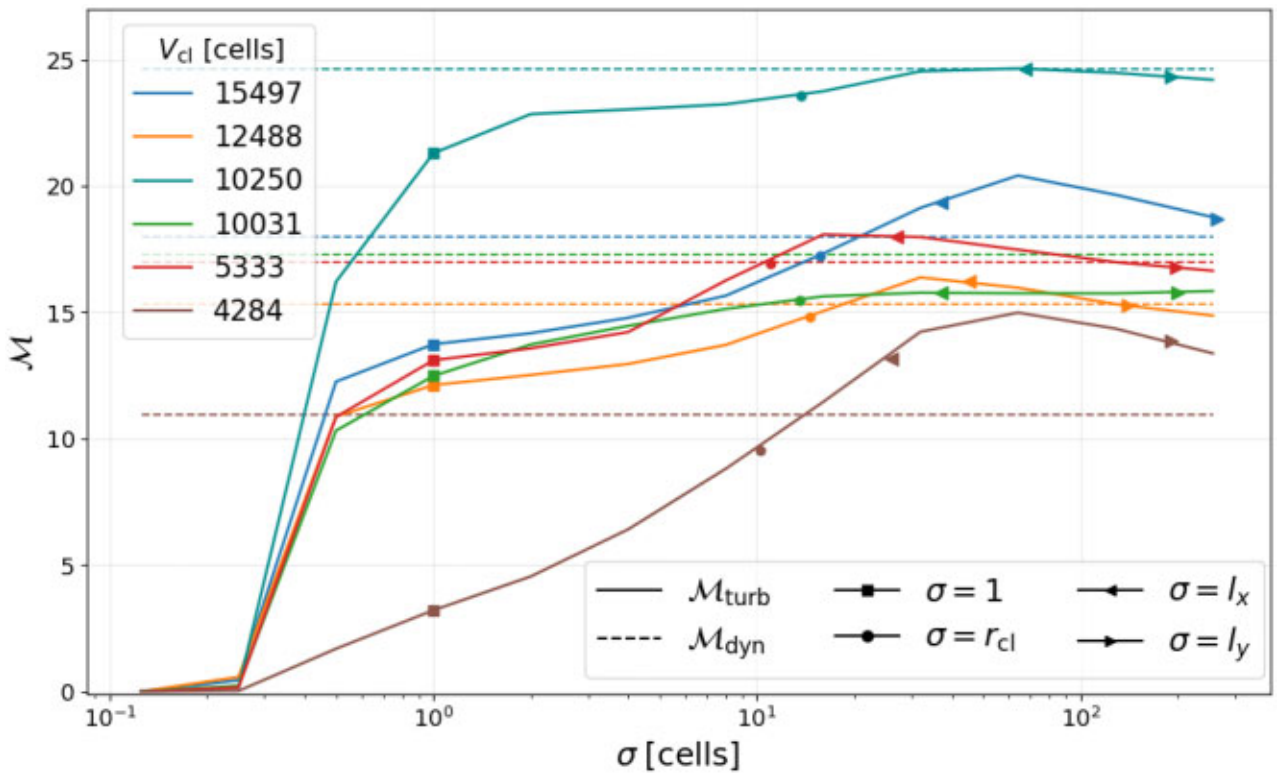

**Figure A1.** Mach number of intercloud motion calculated using two methods. $\mathcal{M}_{turb}$ calculated using a mass-weighted convolution with a Gaussian kernel of width $\sigma$ in solid lines. Compared to $\mathcal{M}_{dyn}$ calculated using the clouds' centre-of-mass velocity. Results shown for $T_{th} = 10^{4.5}$ K at $t = 1.5$ Myr, for the six largest FoF clouds in terms of cloud volume $V_{cl}$, found at that snapshot. With important cloud-size-dependent kernel widths shown in named markers on each curve.

This method, while robust, has a clear dependence on the kernel size $\sigma$ used in the convolution. In order to quantify this dependence, we calculated $\mathcal{M}_{turb}$, for a large span of filtering widths, $\sigma$, on a 100 of the largest clouds found by FoF at multiple snapshots of the low-res simulation. We present the results of this comparison at $t = 1.5$ Myr for six largest clouds at that snapshot in Fig. A1. The values of $\mathcal{M}_{turb}$ as a function of the chosen $\sigma$ are shown in solid curves, while the values of $\mathcal{M}_{dyn}$ presented in dashed horizontal lines. The results shown in the figure are quite typical for other smaller clouds, and different snapshots in the simulation. Additionally, we also show the cloud-dependent convolution kernel widths using specific markers. These are defined as either the maximal length of the cloud in the wind direction $l_y$, or the perpendicular to it, $l_x$. For the largest clouds in our ensemble, $l_x$ reaches $\sim 100$ cells which is comparable to the width of the entire simulation ($128 \times 1920 \times 128$).

As discussed in Section 5, the geometry of the clouds is highly elongated, resembling cold gas 'noodles' stretched along the direction of the wind, that connect with other clouds via narrow filaments of cold gas. Thus, $l_y$ and $l_x$ are more representative of the box that encompasses the whole cloud. A better, though still not accurate, approximation of a typical cloud size would be to assume that it has a spherical geometry and a typical radius $r_{cl} \propto V^{1/3}$, also shown in Fig. A1. The actual width of each 'noodle' might be as narrow as 2–3 cells in the $\hat{x}$- and $\hat{z}$-directions. Filtering on this, resolution-dependent scale, would closely resemble the value of $\mathcal{M}_{turb}$ with $\sigma = 1$ cell.

Fig. A1 can be roughly divided to two regimes. First, for kernel sizes which closely resemble the size of the clouds, $\sigma \sim 1\, r_{cl}$, the turbulent Mach number is smaller than its dynamic counterpart, while still remaining highly supersonic. $\mathcal{M}_{turb} \sim 3 - 20$, as compared to $\mathcal{M}_{dyn} \sim 10 - 25$. For the entire data set of analysed clouds and snapshots, for $\sigma \sim 1\, r_{cl}$, the ratio between the two methods of calculation shows values of $\mathcal{M}_{turb}/\mathcal{M}_{dyn} \sim 0.3 - 0.5$. This is the range of kernel sizes which best represents actual cloud dimensions, and within it the turbulent and cloud edges related motions (such as compression or turbulent mixing layers, Fielding et al. 2020) account for $\sim 50$ per cent of the entirety of the residual velocities within a cloud.

The second part of Fig. A1 is for $\sigma \sim l_x - l_y$. In this area, the values of $\mathcal{M}_{turb}$ rise higher than $\mathcal{M}_{dyn}$. We believe this to be a numeric effect of the convolution, caused by a kernel size that is too large and does not capture the dynamics of a cloud. For such large kernels, edge effects of the simulation box come into play, as well as overlapping and averaging with clouds which are topologically disconnected, but can overlap each other in a wide enough convolution. Also, we assume that for a density contrast between cloud and wind gas of $\eta \gtrsim 10^2$, the mass weighting causes minimal mixing of hot wind velocities into the clouds. This assumption becomes less robust for highly stretched, filamentary geometries of the clouds when convolving with a symmetric 3D kernel. Since the cloud velocity is calculated without filtering out the hot wind, for a stretched infinite filament, the number of cold cells that contribute to the mean velocity calculation is proportional to the smoothing length $\sigma$, but the number of hot cells is proportional to $\sigma^3$, so the contamination ratio scales as $\sigma^2$. Thus, for $\sigma \gtrsim 10$ cells, an equal amount of cold and hot cells is included in the convolution, skewing the velocity field towards the faster hot wind.

Overall, our definition of $\mathcal{M}_{dyn}$ includes additional velocity components besides the commonly used $\mathcal{M}_{turb}$. Regardless of this difference, for a reasonable choice of $\sigma$ both definitions qualitatively agree, and we measure highly supersonic motions within the cold clouds in our wind/multicloud flow. These motions create shocks, which effectively transfer kinetic energy into its thermal counterpart, raising the typical steady-state temperature of the clouds, as well as contributing considerable non-thermal pressure.

This paper has been typeset from a TEX/LATEX file prepared by the author.